\documentclass[preprint,draft,amsmath,amssymb,prd,superscriptaddress]{revtex4}
\newcommand{\Bderiv}{\eth}
\newcommand{\barSigma}{\boldsymbol{\Sigma}}
\newcommand{\barGamma}{\boldsymbol{\Gamma}}
\newcommand{\barn}{{\sf n}{}}
\newcommand{\bara}{{\sf a}{}}
\newcommand{\barb}{{\sf b}{}}
\newcommand{\barc}{{\sf c}{}}
\newcommand{\barh}{{\sf h}{}}
\newcommand{\barq}{{\sf q}{}}
\newcommand{\bark}{{\sf k}{}}
\newcommand{\barM}{{\sf M}{}}
\newcommand{\barD}{{\sf D}{}}
\newcommand{\barsigma}{\boldsymbol{\sigma}}
\newcommand{\half}{{\textstyle\frac{1}{2}}}
\newcommand{\bom}{{}{\sc b}\'o{\sc m}{ }}
\newcommand{\adm}{{}{\sc adm}{ }}
\newcommand{\rt}{{}{\sc rt}{ }}  
\usepackage{graphicx}
\usepackage{dcolumn}
\usepackage{bm}

\begin{document}
\title{Center of mass integral in canonical
general relativity}
\author{D.~Baskaran}
\altaffiliation[Now at ]{Department of Physics and   
Astronomy, University of Wales, Cardiff, CF24 3YB, Wales, UK}
\email{spxdb@astro.cf.ac.uk}
\affiliation{Sternberg Astronomical Institute, Moscow State 
University, Universitet Prospect 13, Moscow 119899, RUSSIA}
\author{S.~R.~Lau}
\email{lau@amath.unc.edu}
\affiliation{Applied Mathematics Group,
Department of Mathematics, University of North 
Carolina, Chapel Hill,
NC 27599-3250 USA}
\author{A.~N.~Petrov}
\email{anpetrov@rol.ru}
\affiliation{Sternberg Astronomical Institute, Moscow State University,
Universitet Prospect 13, Moscow 119899, RUSSIA}
\date{12 September 2003}
\begin{abstract}
For a two-surface $B$ tending to an infinite--radius round 
sphere at spatial infinity, we consider the Brown--York boundary integral 
$H_{B}$ belonging to the energy sector of the gravitational Hamiltonian. 
Assuming that the lapse function behaves as $N \sim 1$ in the limit, we 
find agreement between $H_{B}$ and the total Arnowitt--Deser--Misner energy, 
an agreement first noted by Braden,
Brown, Whiting, and York. However, we argue that 
the Arnowitt--Deser--Misner mass--aspect differs from a gauge invariant 
mass--aspect by a pure divergence on the unit sphere. We also examine the 
boundary integral $H_{B}$ corresponding to the Hamiltonian generator of an 
asymptotic boost, in which case the lapse 
$N \sim x^{k}$ grows like one of the 
asymptotically Cartesian coordinate functions. Such a two--surface integral 
defines the $k$th component of the center of mass for (the initial data 
belonging to) a Cauchy surface $\Sigma$ bounded by $B$. In the large--radius
limit, we find agreement between $H_B$ and an integral introduced by Beig 
and {\sc \'{o}} Murchadha as an improvement upon the center--of--mass 
integral first written down by Regge and Teitelboim. Although both $H_B$
and the Beig--{\sc \'{o}} Murchadha integral are naively divergent, they
are in fact finite modulo the Hamiltonian constraint. Furthermore, we 
examine the relationship between $H_B$ and a certain two--surface integral 
which is linear in the spacetime Riemann curvature tensor. Similar integrals 
featuring the curvature appear in works by Ashtekar \& Hansen, Penrose, 
Goldberg, and Hayward. Within the canonical 3+1 
formalism, we define gravitational energy and center--of--mass as 
certain moments of Riemann curvature.
\end{abstract}
\maketitle
\section{Introduction}
Realization of the Lie algebra $\mathfrak{g}$ of the 
Poincar\'{e} group $G$ as an algebra of differential operators 
features
\begin{equation}
M^{xt} = x \partial/\partial t + t \partial/\partial x 
\label{diffboost}
\end{equation}
as the generator of a pure boost in the $x$--$t$ plane. 
A more remarkable realization of $\mathfrak{g}$ arises in 
classical, canonical, and special relativistic field theory, 
a setting where field integrals play the role of group 
generators and Poisson bracket serves as Lie bracket. The
incarnation of the generator (\ref{diffboost}) in this 
setting is \cite{Schwinger,Weinberg}
\begin{equation}
M^{xt} = - \int_\Sigma d^3x\, x T^{tt}(\mathbf{x},t) + t P^x\, ,
\label{cftboost}
\end{equation}
where $T^{\mu\nu}$ is the Belinfante tensor, the same as 
the stress--energy--momentum tensor for simple theories,
and
\begin{equation}
P^x = \int_\Sigma d^3x\, T^{tx}(\mathbf{x},t)
\end{equation}
is the $x$--component of the canonical field momentum. 
Integration in these expressions is over an inertial 
three--dimensional hyperplane $\Sigma \simeq E^3$ determined 
by fixation of the coordinate time $t$. Notice that the
first term on the {\sc rhs} of Eq.~(\ref{cftboost}), the one 
built with the material energy density $T^{tt}(\mathbf{x},t)$, 
defines (minus) the $x$--component of the field's $\Sigma$ 
center of mass, whence the numerical value for $M^{tx} = 
- M^{xt}$ obtained via evaluation of the integrals is 
generally not equal to this center--of--mass component. 
Via the Poisson bracket $M^{xt}$ generates the change 
$\{\Psi (\mathbf{x},t), M^{xt}\}$ in the classical field 
$\Psi (\mathbf{x},t)$ corresponding to the infinitesimal 
$x$--$t$ boost, and for points $\mathbf{x}$ close to the 
origin this change is chiefly governed by $t P^x$ in 
(\ref{cftboost}). So indeed this term must be present.

One encounters a somewhat different situation in canonical 
general relativity ({\sc gr}) and the arena of spacetimes 
asymptotically flat at spatial infinity. Although in general 
such spacetimes possess no group of isometries, one can 
nevertheless realize $G$ as an asymptotic isometry group by 
writing down gravitational Hamiltonians which generate the 
asymptotic symmetries. \cite{ReggeTeitelboim,BeigOMurchadha}
We draw attention to a salient feature of the resulting description. 
Namely, the generator of an asymptotic boost ---say the one 
corresponding to $M^{tx}$ above--- has a form incorporating only 
the first term on the {\sc rhs} of Eq.~(\ref{diffboost}), or 
Eq.~(\ref{cftboost}) for that matter. (These remarks pertain 
only to boost generators in {\sc gr}, as the remaining 
generators of asymptotic translations and rotations look 
just like they should.) Moreover, the numerical value for 
the $M^{tx}$ generator in {\sc gr} equals the 
$x$--component of the gravitational field's center of mass. 
Rather symbolically, the {\em value} of the said generator 
takes the form 
\begin{equation}
M^{tx} = \int_B d^2x \sqrt{\sigma} x \tau^{tt}\, ,
\label{grboost}
\end{equation}
where $B$ is a large two--surface with spherical topology,
$d^2x\sqrt{\sigma}$ is the proper area element of $B$, and 
$\tau^{tt}$ is an energy density stemming from 
a ``boundary stress--energy--momentum tensor'' 
$\tau^{\alpha\beta}$. \cite{BrownYork}

The following heuristic argument sheds light on this salient 
feature. For the scenario of asymptotic flatness towards spatial 
infinity, ``asymptotic'' crudely means fixed--time and {\em 
arbitrarily large} radial separation from the ``origin,'' perhaps 
actually a closed two--surface with spherical topology. In $E^3$ 
such arbitrarily large radial separation corresponds ---except 
on a set of measure zero, the $y$--$z$ plane--- to large values of 
$x$; therefore, the asymptotic modifier roughly indicates 
that $x \partial/\partial t$ is the dominant term in 
Eq.~(\ref{diffboost}). The ramifications of these simple 
observations for {\sc gr} stem from the following key point: 
the numerical value (determined by evaluation on a classical
solution) of a canonical generator in {\sc gr} is a surface 
integral at infinity, that is to say a boundary integral like the 
one in Eq.~(\ref{grboost}) over a two--surface surface $B$ 
enclosing the ``origin'' and whose points are uniformly separated 
from it by an arbitrarily large radial distance. Therefore, the
integration in Eq.~(\ref{grboost}) is quite unlike the integration 
in Eq.~(\ref{cftboost}) in the following sense. The subset of $B$
on which $x$ is small is itself {\em arbitrarily} small 
(essentially just a ``great circle'' around $B$), 
suggesting that the expression (\ref{grboost}) ---analogous 
to (minus) $x \partial/\partial t$ only--- is permissible. In the end, of 
course, a test for whether one has chosen the correct 
boost generator is whether it serves its part in a consistent 
representation of $\mathfrak{g}$ under the Dirac algebra 
determined by the Poisson bracket. The asymptotic boost generator 
we consider below has long since measured up on this 
count. \cite{ReggeTeitelboim,BeigOMurchadha} 
Our simple discussion here is meant only to draw attention to 
the discussed feature, one seemingly neglected in the literature. 
However, we do point out that within the framework of the 
Lagrangian (rather than Hamiltonian) field formulation of {\sc gr}, 
Ref.~\cite{Petrov95} considered integrals of motion at spatial 
infinity which were in the spirit of Eq.~(\ref{cftboost}) and special 
relativistic theory, while (on classical solutions) their numerical 
values would be of the form Eq.~(\ref{grboost}). Since a Lagrangian 
approach does not select preferred $\Sigma$ hyperplanes, both types of 
integrals would thus be treated on the same footing. 

Brown and York have written down a geometric expression for 
the integral appearing in Eq.~(\ref{grboost}). They consider 
the following boundary term belonging to the ``energy sector'' of 
the gravitational Hamiltonian: \cite{BrownYork,BBWY}
\begin{equation}
       H_{B} = \frac{1}{8\pi}\int_{B} d^{2}x 
               \sqrt{\sigma}\, N
       \big(k - k|^{\rm ref}\big)\, ,   
\label{BYintegral}
\end{equation}
with $B$ and $d^{2}x\sqrt{\sigma}$ as before. Note that $B$ is 
(perhaps one element of) the boundary $\partial\Sigma$ of a
hypersurface $\Sigma$. $N$ is a smearing lapse function, $k$ is 
the mean curvature associated with the embedding of $B$ in $\Sigma$, 
and  $k|^{\rm ref}$ is the reference mean curvature of $B$ 
associated with an isometric embedding of $B$ in an auxiliary 
Euclidean three-space $\barSigma \simeq E^3$. That is to 
say, whether $B$ is viewed as a surface in $\Sigma$ or in 
$\barSigma$, it has the same two-metric $\sigma_{ab}$. 
Hence, the integral (\ref{BYintegral}) is essentially the 
difference of the total mean curvatures for the two embeddings. As 
shown in \cite{BrownYork}, such a surface integral must be 
added to the smeared Hamiltonian constraint
\begin{equation}
H_{\Sigma} = \int_{\Sigma}{\rm d}^{3}x N {\cal H}
\, ,
\label{smearedHam}
\end{equation}
in order to obtain a $\Sigma$ hypersurface functional 
$H = H_{\Sigma} + H_{B}$ which is differentiable on the 
standard gravitational phase space. The $H$ functional 
is differentiable provided $k|^{\rm ref}$ is determined solely 
by the $B$ metric. Other choices for $k|^{\rm ref}$ are
possible \cite{BLYss,Lau2}, but except for one passing remark 
will not be considered here. In 
appropriate limiting scenarios it is known that $H_B$ agrees 
with the standard Arnowitt--Deser--Misner ({\sc adm}) 
notion \cite{ADM} of total energy for spacetimes asymptotically 
flat at spatial infinity, the Trautman--Bondi--Sachs 
notion \cite{TBS} of total energy--momentum for spacetimes 
asymptotically flat at null infinity, and the Abbott--Deser 
notion \cite{AbbottDeser} of conserved mass for spacetimes 
asymptotically anti--de Sitter at 
infinity. \cite{HawkingHorowitz,BLY1,Brown,Lau2}

In this paper we consider the the scenario of asymptotic 
flatness towards spatial infinity ({\sc spi}), in which case 
the $\Sigma$ gravitational initial data $(h_{ij}, K_{ij})$, 
spatial metric and extrinsic curvature, obey certain fall-off 
conditions specified below. If we adopt this setting and assume 
that $B$ tends to an infinite--radius round sphere at 
{\sc spi}, then we may obtain physical characterizations of the 
initial data in terms of the integral (\ref{BYintegral}). For 
instance, if the lapse obeys $N \sim 1$ in the said limit, 
then $H$ generates a pure time translation asymptotically, and 
the ``on--shell'' (meaning on solutions to ${\cal H} = 0$) value 
$E_{\infty} \equiv \lim_{r\rightarrow\infty}
H_{B}$ of $H$ defines the total $\Sigma$ energy. 
Braden et al.~\cite{BBWY} and later
Hawking and Horowitz \cite{HawkingHorowitz} have noted
that such a definition of energy agrees with the 
standard {\sc adm} notion of total energy, although here we 
establish this equivalence in much more detail. We argue that
the Arnowitt--Deser--Misner mass--aspect differs from a gauge 
invariant mass--aspect by a pure divergence on the unit sphere.

Moreover, if the lapse grows like $N \sim x^{k}$, where $x^{k}$ is 
one of the Cartesian coordinate functions near {\sc spi}, then $H$ 
generates an asymptotic boost, and the on--shell value 
$M^{\perp k}_\infty \equiv \lim_{r\rightarrow \infty} H_{B}$ of $H$ 
defines the $k$th component of the $\Sigma$ center of 
mass. \cite{ReggeTeitelboim,BeigOMurchadha} 
For this asymptotic lapse behavior, we find agreement between 
$M^{\perp k}_\infty$ and the integral introduced by Beig and {\sc \'{o}} 
Murchadha in Ref.~\cite{BeigOMurchadha} (\bom hereafter) as an 
improvement upon the center of mass integral first written down 
by Regge and Teitelboim in Ref.~\cite{ReggeTeitelboim} 
(\rt hereafter). This is our first main result. We also establish
the relationship between $H_B$ and a certain two--surface integral 
which is linear in the spacetime Riemann curvature tensor. This is
our second main result. Similar two--surface integrals featuring 
the curvature appear in works on gravitational energy--momentum by 
Ashtekar \& Hansen \cite{AshtekarHansen}, Penrose \cite{Penrose}, 
Goldberg \cite{Goldberg0}, and S.~Hayward \cite{SHayward}, among others.
We believe our results to be of relevance for comparison 
between the standard 3+1 approach to spatial infinity and 
more formal treatments based on compactification 
arguments. \cite{Ashtekar,BeigOMurchadha} The results of this work
complement those given in the seminal 
Refs.~\cite{ReggeTeitelboim,BeigOMurchadha}, as well as those given 
in Ref.~\cite{BLY3} which mentioned that the present work would
appear.

\section{Preliminaries}
\subsection{Fall--off for metric and extrinsic curvature}
Recall the definition of a spacetime which is asymptotically 
flat towards spatial infinity given by {\sc b}\'{o}{\sc m}. Such 
a spacetime possesses spatial sections on which there are so--called 
asymptotically Cartesian coordinates $x^k$ with corresponding polar 
coordinates $(r,\theta,\phi)$. Further, with respect to these 
coordinates, the large--$r$ perturbations $\delta h_{ij}$ and 
$\delta K_{ij}$ of the $\Sigma$ three-metric and extrinsic 
curvature tensor are defined by
\begin{subequations}\label{BOMzero}
\begin{eqnarray}
h_{ij} & = & f_{ij}+\delta h_{ij}
\sim  f_{ij} + a_{ij} (\nu^k)r^{-1} + 
b_{ij}r^{-1-\varepsilon}
\label{BOMzero1} \\
K_{ij} & = & 0_{ij} + \delta K_{ij} \sim
d_{ij}(\nu^k) r^{-2} + e_{ij}
r^{-2-\varepsilon}\, ,
\label{BOMzero2}
\end{eqnarray}
\end{subequations}
where $0 <\varepsilon \le 1$. In 
(\ref{BOMzero1}) the flat metric $f_{ij}$ is ${\rm diag} (1,1,1)$, 
$\nu^k = x^k/r$, the $O(1)$ function $a_{ij}$ is of even parity 
[that is to say, $a_{ij}(-\nu^k) = a_{ij}(\nu^k)$], and the
$O(1)$ angular function $b_{ij}$ is of undetermined parity. 
In (\ref{BOMzero2}) $0_{ij}$ is the zero tensor, $d_{ij}$ is a 
function of odd parity [that is to say, $d_{ij}(-\nu^k) = 
-d_{ij}(\nu^k)$], and the $O(1)$ angular function $e_{ij}$ is of 
undetermined parity. The three-metric expansion considered by \rt 
is of the same form, but with $\varepsilon = 1$ in (\ref{BOMzero1})
and (\ref{BOMzero2}).
Cartesian differentiation of (\ref{BOMzero}) yields behavior obtained 
via term--by--term multiplication by $r^{-1}$ and parity reversal on 
leading terms. Considering (\ref{BOMzero1}), one can say that 
asymptotically Cartesian coordinates define their own Euclidean
background $E^3$ at {\sc spi}, with respect to which the perturbations 
$\delta h_{ij}$ are defined.

In this paper we work with the expansion
\begin{equation}
h_{ij} = f_{ij}+\delta h_{ij}
\sim  f_{ij} + a_{ij} (\nu^k)r^{-1} +
b_{ij}r^{-1-\varepsilon} + c_{ij}r^{-2}\, ,
\label{BOM}
\end{equation}
with $c_{ij}$ also an $O(1)$ angular function of 
undetermined parity. Hence, we are allowing for the possibility
of a {\em single} power, say for example $r^{-3/2}$, lying 
between $r^{-1}$ and $r^{-2}$. We could of course
adopt a more general expansion, say
\begin{equation}
h_{ij} = f_{ij}+\delta h_{ij}
\sim  f_{ij} + a_{ij} (\nu^k)r^{-1} + \sum_{q=1}^{N}
b^{(q)}_{ij}r^{-1-\varepsilon_q} + c_{ij} 
r^{-2}\, ,
\label{BOMnotused}    
\end{equation}
with up to $N$ terms lying between the $r^{-1}$ and $r^{-2}$ orders.
In this case we would have $0 < \varepsilon_1 < \varepsilon_2 < 
\cdots < \varepsilon_N  \leq 1$
and $b^{(q)}_{ij}$ of undetermined parity for each $q$. However,
adoption of such a more general expansion needlessly complicates 
the analysis. Furthermore, we believe that all of our main results 
are also valid for the expansion (\ref{BOMnotused}).

Let us comment on the meaning of the asymptotic symbol $\sim$ in 
this paper. In the asymptotic expansions we write down, if the 
last term written is $O(r^p)$ with $p$ some integer, then the 
next unwritten term is $O(r^{p-\gamma})$ where $0 < \gamma \leq 1$. 
We have switched from $\varepsilon$ to $\gamma$ now in order 
to emphasize following. For the next unwritten $O(r^{-2-\gamma})$ 
term in an equation like (\ref{BOM}), the $\gamma$ need not be 
the same as the $\varepsilon$ in the third term on the {\sc rhs}. 
Furthermore, in an equation such as 
(\ref{BOMzero1}) the $\sim$ indicates that the next unwritten 
term is $O(r^{-2})$, that is to say the next integral power of 
$1/r$ following $r^{-1-\varepsilon}$. Sometimes we include the 
order symbol $O$ and write equalities for the sake of clarity.

\subsection{\bom integral and asymptotic scenarios}
Beig and {\sc \'{o}} Murchadha \cite{BeigOMurchadha}
define a center of mass integral
associated with the Hamiltonian generator of an asymptotic
boost. Namely,
\begin{equation}
M^{\perp}_\infty (N)  =  \frac{1}{16\pi}
       \oint_{\infty}d^2x\sqrt{\sigma}n^k
       \left[N h^{ij}(\partial_{j} h_{ik}
     - \partial_{k} h_{ij}) + 
       (h^{l}{}_ kh^{ij}
     -  h^{i}{}_k h^{lj})
        (h_{ij} - f_{ij})\partial_l N
     \right]\, ,
     \label{BOMintegral}
\end{equation}
where $\oint_{\infty}$ is shorthand for 
$\lim_{r\rightarrow\infty}\int_{B}$ with $B$ a level--$r$ 
two--surface and $n^k$ its outward--pointing normal. Furthermore, here, 
as it should for a boost \cite{ReggeTeitelboim}, the lapse behaves as
\begin{equation}
N =\beta^\bot{}_k x^{k} + \alpha^{-} + O(r^{-\varepsilon})\, ,
\label{N}
\end{equation}
with the $\beta^\bot{}_k = -\beta_k{}^\bot$ constants and
$\alpha^{-}$ an $O(1)$ angular function of odd parity. The 
expression above (\ref{BOMintegral}) is written in terms of asymptotically 
Cartesian coordinates and ordinary partial derivatives, but it is easy to 
render the expression covariant.

Were the 
lapse to go only as $N = 1 + O(r^{-\varepsilon})$, then the \bom 
integral would be the \adm energy. Indeed, one recognizes the first 
term in the integrand (\ref{BOMintegral}) as the familiar \adm 
expression,
\begin{equation}
E_{\infty} = \frac{1}{16\pi}
       \oint_{\infty} d^2x \sqrt{\sigma} 
       n^kh^{ij}(\partial_{j} h_{ik}
     - \partial_{k} h_{ij})\, .
\label{ADMenergy}
\end{equation}
We examine two scenarios in this paper; these being the
{\em energy scenario}, in which the lapse behaves as $N\sim 1$, and 
the {\em center--of--mass scenario}, in which the lapse behaves as 
in Eq.~(\ref{N}). 

We note that the integral (\ref{BOMintegral}) is naively
divergent, although its actual finiteness is ensured by the particular
choice (\ref{N}) of lapse $N$, the even parity of $a_{ij}$ built into 
the asymptotic $\Sigma$ metric (\ref{BOMzero1}), and on account of
the integral $H_\Sigma$ of the corresponding constraint ${\cal H}$. As 
Beig and \'{o} Murchadha show in Appendix C of their 
Ref.~\cite{BeigOMurchadha},
addition of $M^{\perp}_\infty(N)$ to $H_{\Sigma}$ given in 
(\ref{smearedHam}) yields a total expression which is explicitly finite. 
(We also analyze the integral $H_\Sigma$ in our {\em Appendix C},
isolating its divergent contribution via a method different than the
\bom one.)
Therefore, the \bom integral is finite on--shell, that is to say, 
given the vanishing of the Hamiltonian constraint $\cal H$ and in turn 
the volume term $H_\Sigma$. Let us use $M^\perp_B(N)$ to denote the 
integral corresponding to (\ref{BOMintegral}) {\em before} the 
$r\rightarrow\infty$ limit is taken. The asymptotic expansion 
\begin{equation}
M^{\perp}_B(N) \sim\,{}^{-1}M^{\perp}_B r +\, 
                   {}^{(-1+\varepsilon)}
                   M^{\perp}_B r^{1-\varepsilon} 
                   +\, {}^0 M^{\perp}_B r^0
\label{Mperpexpansion}
\end{equation}
for the integral (corresponding now to a large but finite 
$B$ two--surface) particularly elucidates the issues at hand. The
parity conditions built into Eqs.~(\ref{BOMzero1},\ref{N}) 
automatically ensure that the constant ${}^{-1}M^{\perp}_B$ 
in fact vanishes. However, invocation of the 
Hamiltonian constraint is required to ensure that the coefficient 
${}^{(-1+\varepsilon)} M^{\perp}_B$ vanishes. With \rt fall-off the 
integral (\ref{BOMintegral}) is explicitly finite without appeal to the 
Hamiltonian constraint, as the $r^{1-\varepsilon}$ order in the
(\ref{Mperpexpansion}) is absent. Whatever the specified fall--off 
is, in the end $M^\perp_\infty (N) = {}^0 M^\perp_B(N)$.

\subsection{Curvature integral}
The other aforementioned two--surface integral, written in terms of the
Riemann tensor, we wish to consider is the following: 
\begin{equation}
M_{\rm \Re} \equiv
       \frac{1}{16\pi}\sqrt{\frac{A}{4\pi}}
       \int_{B}d^{2}x
       \sqrt{\sigma} N\sigma^{\mu\nu}\sigma^{\lambda\kappa}
       \Re_{\mu\lambda\nu\kappa}\, .
\label{mainlimit0}
\end{equation}
In this equation $\Re_{\mu\nu\lambda\kappa}$ is the spacetime Riemann 
tensor, $A$ is the area of $B$, and the square root factor outside of 
the integral is asymptotic to a single power of the coordinate radius 
$r$. Furthermore, the spacetime representation of the $B$ two--metric
$\sigma^{\mu\nu}$ serves here to project free indices into $B$. 
With the future--pointing normal of $\Sigma$ in spacetime denoted by 
$u^\mu$ and the out--pointing normal of $B$ in $\Sigma$ denoted by 
$n^\mu$, we have $\sigma^{\mu\nu} = g^{\mu\nu} + u^\mu u^\nu - n^\mu n^\nu$. 
As for a physical interpretation of the integral (\ref{mainlimit0}),
consider timelike geodesics whose tangents at points of the hypersurface
$\Sigma$ are given by $u^\mu$. In the absence
of matter $\sigma^{\mu\nu}\sigma^{\lambda\kappa}\Re_{\mu\lambda\nu\kappa}
= -2u^\mu n^\lambda u^\nu n^\kappa \Re_{\mu\lambda\nu\kappa}$, and the
latter quantity controls the rate of change of the vector joining any two 
nearby geodesics and describing the radial displacement between 
them. Hence, the integrand in (\ref{mainlimit0}) dictates the radial
component of the geodesic deviation vector, as one would expect for a
``mass--aspect.'' \cite{Ashtekar} We show below that $H_B$ and $M_\Re$ 
also agree in the $r\rightarrow\infty$ {\sc spi} limit, again for both 
energy and center--of--mass scenarios. The on--shell finiteness of both 
$H_B$ and $M^\perp_B$ as $r \rightarrow \infty$ in the 
weaker \bom fall--off setting is intimately tied to the fact that these 
are correct surface integrals to add to (\ref{smearedHam}), thereby
achieving a functionally differentiable total Hamiltonian $H = H_\Sigma +
H_B$. On the other hand, $M_\Re$ in (\ref{mainlimit0}) has no 
particular relation to the canonical 3+1 Hamiltonian; therefore, 
its on--shell finiteness is far from obvious. See 
Ref.~\cite{SHayward} for a discussion of how an integral very 
similar to $M_\Re$ stems from a dual--null $2+2$ Hamiltonian 
description of general relativity.

Our comparison of $H_B$ with $M_\Re$ leads to the following alternative
definitions for energy and center of mass for gravitational
initial data sets. For the energy we find
\begin{equation}
E_\infty = \frac{1}{16\pi} \oint d\Omega\,
    {}^3\!\!\left[\sigma^{\mu\nu}\sigma^{\lambda\kappa}
    \Re_{\mu\lambda\nu\kappa}
\right]\, ,
\label{averageR3}
\end{equation}
where the superscript $3$ means take the coefficient of the
$O(r^{-3})$ term in the radial expansion of the curvature component 
$\sigma^{\mu\nu}\sigma^{\lambda\kappa}\Re_{\mu\lambda\nu\kappa}$ and
here the averaging is over the unit sphere.
This definition may be easily generalized to a ``superenergy'' ---the
charge integral corresponding to a general time supertranslation at 
{\sc spi}--- by placing an angle--dependent smearing function in front 
of the mass--aspect ${}^3\![\sigma^{\mu\nu}\sigma^{\lambda\kappa}
\Re_{\mu\lambda\nu\kappa}]/4$, in other words using 
$N = O(1)$ rather than $N=1$ in Eq.~(\ref{averageR3}). We stress 
that the {\sc adm} mass--aspect (determined from the integrand lifted 
from the expression for the {\sc adm} energy) differs from this manifestly 
gauge--invariant mass--aspect by a pure divergence on the unit sphere.
See the concluding remarks in {\em Section V} for further details.
Arguably, the {\sc adm} mass--aspect does not define a valid superenergy, 
since such a definition not would generally vanish for Minkowski 
spacetime. For the coordinates $M^{\perp k}_\infty$ of the data 
set's center of mass, we 
find
\begin{equation}
M^{\perp k}_\infty 
  = \frac{1}{16\pi} \oint d\Omega\,
    {}^4\!\!\left[\sigma^{\mu\nu}\sigma^{\lambda\kappa}
    \Re_{\mu\lambda\nu\kappa}
\right]\nu^k\, ,
\label{prelimMperpk}
\end{equation}
with a similar meaning for the superscript $4$.

\section{$H_B$ and the Beig--\'{o} Murchadha integral}
In this section we consider the $H_B$ integral (\ref{BYintegral}), with 
$B$ a level--$r$ surface, showing that its $r\rightarrow\infty$ limit 
coincides with the \bom integral (\ref{BOMintegral}). To set up and compute 
the limit, we shall introduce a background metric $\barh_{ij}$ on $\Sigma$ 
via a certain embedding of $B$ into an auxiliary Euclidean three--space 
$\barSigma \simeq E^3$. All quantities associated with the background will
be denoted by either san serif or boldface letters. In particular, we use 
$\bark$ to represent $k |^{\rm ref}$, or more precisely an approximation 
to $k |^{\rm ref}$ of sufficient accuracy to compute the limit.

\subsection{Key identities}
Let us first lay some groundwork necessary to obtain useful identities 
for both $k-{\bark}$ and $N(k-{\bark})$. Assume that we have a 
three--dimensional slice $\Sigma$, equipped with two distinct 
proper Riemannian three--metrics: $h_{ij}$ with compatible covariant 
derivative $D_i$ and $\barh_{ij}$ with compatible covariant derivative 
$\barD_i$. Viewing $h_{ij}$ as the physical metric and $\barh_{ij}$ 
as the background metric, we set
\begin{equation}
h_{ik} = \barh_{ik} + \Delta h_{ik} \, ,
\qquad h^{ik} = \barh^{ik} + \Delta q^{ik}\, ,
\label{perturbations}
\end{equation}
where the terms $\Delta h_{ik}$ and $\Delta q^{ik}$ contain {\em all}
orders of perturbations.

Assume that a radial coordinate $s$ foliates $\Sigma$ into a nested 
family of smooth closed two--surfaces. Later, $s$ will be the radial
coordinate $r = (x_k x^k)^{-1/2}$ determined by asymptotically
Cartesian coordinates $x^k$. We use $s$ at this point to emphasize
the fact that here our calculations are not tied to a particular type 
of two--surface. Loosely, we use the letter $B$ to
represent both the $s$--foliation of $\Sigma$ itself and a particular 
slice (or leaf) of the foliation determined by setting $s$ equal to a 
constant value $s_{0}$. 
Respectively, let $\sigma_{ab}$ and $\barsigma_{ab}$ denote the
metrics induced on a generic level--$s$ slice by the $h_{ij}$ and
$\barh_{ij}$ metrics. Our central assumption is that the metrics 
$\sigma_{ab}$ and $\barsigma_{ab}$ agree in a sense made precise below.
The background metric introduced in (\ref{perturbations}) will not 
be arbitrary, rather it will be defined by an (essentially) isometric
embedding of $B$ into Euclidean space and as such will be the flat metric
belonging to a Euclidean three--space $\barSigma \simeq E^3$.

Let $n_k = M \partial_k s$
be the outward-pointing normal covector for $B$ as a submanifold of the
Riemannian space $(\Sigma\,,\,h_{ij})$. Likewise, let $\barn_k =
\barM \partial_k s$ be the outward-pointing normal covector for $B$ as
a submanifold of the Riemannian space $(\Sigma\, ,\,\barh_{ij})$. 
The function $M = [h^{ij} (\partial_i s)(\partial_j s)]^{-1/2}$ ensures
that $n_k n^k = 1$, and likewise $\barM$ ensures the $\barn_k \barn^k =
1$. Above and in what follows, indices on physical objects are
lowered and raised with $h_{ij}$ and 
its inverse $h^{ij}$, whereas indices on background
objects are lowered and raised with $\barh_{ij}$ and its inverse
$\barh^{ij}$. Again, we let $\Delta n^{i} = n^i - \barn^i$ stand for all
orders of perturbations. In $\Sigma$ coordinates the induced (projection)
metrics for the $s$ foliation of our two Riemannian spaces are
\begin{equation}
\sigma^{ik} = h^{ik} - n^i n^k\, , \qquad
\barsigma^{ik} = \barh^{ik} - \barn^i \barn^k\, .
\label{projections}
\end{equation}
Now our requirement that the metrics induced on $B$ by $h_{ij}$ and
$\barh_{ij}$ agree can be rewritten as 
\begin{equation}
\Delta\sigma^{ij} = 0\, ,
\label{twoagree}
\end{equation}
where $\Delta\sigma^{ik} = \sigma^{ik} - \barsigma^{ik}$. Actually, 
this condition is too strong. Below, when $s$ is $r = (x_k x^k)^{1/2}$ 
we shall only demand that $\Delta\sigma^{ij} = O(r^{-2-\varepsilon})$. 
However, as we shall retain $\Delta\sigma^{ij}$ in our calculations, 
there is no harm in considering this stronger agreement for the 
time--being, and doing so more clearly demonstrates the reason for 
working with the inverse two--metric. We stress that
while working with {\em three dimensional}
indices, one must enforce equality of the
inverse two--metrics as above in order to ensure that $\sigma^{ab} =
\barsigma^{ab}$ on $B$ (the same, of course,  as $\sigma_{ab}$ and
$\barsigma_{ab}$ agreeing on $B$). Indeed, if we choose $s$ as the first
coordinate $x^1$, then the index $a$ on the $B$ coordinate $x^a$ runs
over $2,3$ and we have
\begin{equation}
\sigma^{1k} = \barsigma^{1k} = 0\, .
\label{nsigiszero}
\end{equation}
That is to say, $\sigma^{ik}$ and $\barsigma^{ik}$ have only $2,3$
components (i.~e.~$B$ components). In contrast, $\sigma_{1k} \neq 0$,
$\barsigma_{1k} \neq 0$ and $\sigma_{1k} \neq \barsigma_{1k}$ in general
on $B$. Therefore, $\sigma_{ik}$ and $\barsigma_{ik}$ need not agree on
$B$. In other words, the equation with lower {\em three dimensional}
indices which is analogous to Eq.~(\ref{twoagree}) is generally not
valid, although, of course,  $\sigma_{ab} = \barsigma_{ab}$ on $B$.

With the assumptions spelled out in the preceding paragraph, we now 
collect the promised identities. Let $N$ be a smearing function and
\begin{equation}
k = - D_i n^i\, , \qquad
\bark = -\barD_i \barn^i\, .
\end{equation}
We then have the following identities:
\begin{subequations}\label{identity}
\begin{eqnarray}
k - \bark
             & =  &               \half
                                  \barn^i h^{kl}\left(\barD_k
                                  \Delta h_{il}
                                - \barD_i \Delta h_{kl}\right)
                                + \barn_{k} \barD_l
                                  \left(\barn^{[l}\Delta n^{k]}\right)
                     \label{identity1} \\
             &   & + \half \barn^i \Delta h_{il}\barD_k \Delta q^{kl}
                   + \half \barn_i \barD_k \left(\Delta n^i\Delta n^k
                     \right)
                     \nonumber \\
             &   & - \half h^{kl}\Delta n^i \barD_i \Delta h_{kl}
                   - \half \Delta \sigma^{ik} \barD_i \barn_k 
                     \, , \nonumber \\
           &   &           \nonumber \\
N(k-\bark) & = &            \half N\barn^i h^{kl}\left(\barD_k
                               \Delta h_{il}
                             - \barD_i \Delta h_{kl}\right)
                             - \half\left(\barn^k h^{il}\Delta h_{ik}
                             - \barn^l h^{ik}\Delta h_{ik}\right)
                               \barD_{l} N \label{identity2} \\
             & & + \barn_{k} \barD_l \left(N\barn^{[l}\Delta n^{k]}\right)
                 + \half N \barn^i \Delta h_{il} \barD_k \Delta q^{kl}
                 + \half N \barn_i \barD_k \left(\Delta n^i\Delta n^k
                   \right)
                   \nonumber \\
             & & - \half N h^{kl}\Delta n^i \barD_i \Delta h_{kl}
                 + \half\left(\barn^l \barh_{ik}\Delta n^i \Delta n^k
                 - \barn_i \Delta n^i \Delta n^l\right)\barD_l N
                   \nonumber \\
             & & - \half N\Delta \sigma^{ik} \barD_i \barn_k
                 + \half \barn^l \barh_{ik} \Delta\sigma^{ik} \barD_l N 
                   \, ,\nonumber
\end{eqnarray}
\end{subequations}
with the first identity following from the second upon assuming that $N$ 
is constant and unity. We derive these identities in {\em Appendix A.1}.
Note that $\Delta \sigma^{ij}$ could be replaced in these identities via
use of the appendix Eq.~(\ref{idone}).

\subsection{Construction of $\barh_{ij}$ and various coordinates}
We construct a diffeomorphism between $\Sigma$ and $\barSigma 
\simeq E^3$ as follows. Take a level--$s$ surface $B$ in $\Sigma$, say 
the one determined by $s = s_0$, and embed it in  $\barSigma$. At this 
point we
make no assumption that this embedding is isometric. In $\barSigma$
assume that $B$ is also a level--$s$ coordinate surface of value $s_0$.
Label the points on this level surface in $\barSigma$ by their 
coordinate values $x^a$ inherited from $B$ in $\Sigma$. Now extend the
the coordinates off of $B$ to a system $(s,x^a)$ on $\barSigma$ in a
region surrounding $B$. One way of doing this would be to construct
Riemann normal coordinates. The construction described gives us a
diffeomorphism, as $(s,x^a)$ label points in both $\Sigma$ and
$\barSigma$. Further, this diffeomorphism identifies level--$s$ 
surfaces in $\Sigma$ with corresponding ones in $\barSigma$, providing
us with the set--up in {\em Section III.A} where we can work with 
a single $\Sigma$ equipped with two distinct proper Riemannian
three--metrics. We stress that with this construction $\barh_{ij}$ 
is a flat Euclidean metric, although it need not be the trivial metric
diag$(1,1,1)$. Let us make a few comments here meant to highlight 
the exceptional nature of the foregoing construction that occurs 
when $s$ is the radius $r = (x_k x^k)^{1/2}$ stemming from 
asymptotically Cartesian coordinates. Suppose that the embedding 
into $\barSigma$ of the original two--surface $B$ defined by $s = s_0$ 
were an exact isometry, which we can guarantee via rather mild 
assumptions on the Ricci scalar $\cal R$ of $B$. Then one would 
not expect that level--$s$ surfaces in $\barSigma$ neighboring 
this initial surface would also be isometric to their counterparts 
in $\Sigma$. Indeed, were this the case, one would have a 
foliation of (an annulus of) flat Euclidean space in which an 
infinite number of slices were exactly isometric to slices 
belonging to a foliation of the non--trivial Riemannian 
space $(\Sigma, h_{ij})$. This would seem to us an overly 
restrictive situation to achieve. However, while working with 
the coordinate $r$, we shall find it possible to nearly achieve 
this situation by relaxing the requirement that the isometries 
are exact, instead assuming that they hold approximately through 
some appropriate order in the small parameter $1/r$. Even subject to this
relaxation, $\bark$ and $k |^{\rm ref}$ will agree to an accuracy 
sufficient to compute the $r\rightarrow\infty$ limit of 
(\ref{BYintegral}) with $N(k-\bark)$ in place of $N(k-k|^{\rm ref})$.

Equation (\ref{perturbations}) has been viewed in the system 
$(s,x^a)$ of coordinates just discussed,
\begin{equation}
h_{ij}(s,x^a) = \barh_{ij}(s,x^a) + \Delta h_{ij}(s,x^a)\, .   
\label{first}
\end{equation}
The splitting above depends on (i) the initial choice of $B$ 
two--surface through the embedding of $B$ in $\barSigma$ and on (ii) 
how the coordinates are extended off of $B$ once the embedding is 
carried out. (Let us just loosely say that the splitting 
depends on the $B$ embedding.) Nevertheless, our calculations are 
covariant on $\Sigma$, and we can go to any other arbitrary 
coordinates. For example, from the system $(s,x^a)$ on $\barSigma$, 
we may transform to a truly Cartesian system $X^k$ on $\barSigma$. 
Via the constructed diffeomorphism, the system $X^k$ may also be 
placed on $\Sigma$. Since by definition coordinate transformation 
to the system $X^k$ makes $\barh_{ij}$ the Kronecker delta (we 
denote this diagonal flat--space metric by $f_{ij}$), adoption of
the system $X^k$ on $\Sigma$ yields the splitting 
\begin{equation}
h_{ij}(X^k) = f_{ij} + \Delta h_{ij}(X^k)\, ,
\label{y-barSig}
\end{equation}
which is quite similar to the type of decomposition used by  \rt 
and  \bom in the system $x^k$ of coordinates, namely 
Eq.~(\ref{BOM}).        

Now assume that the coordinate $s$ is in fact the radius $r$
stemming from asymptotically Cartesian coordinates $x^k$. Then 
we may write Eq.~(\ref{first}) as
\begin{equation}
h_{ij}(r,\theta,\phi) = \barh_{ij}(r,\theta,\phi) 
                      + \Delta h_{ij}(r,\theta,\phi)\,
.
\label{firstprime}
\end{equation}
As mentioned, now the situation will be that all {\em large} 
level--$r$ surfaces will have essentially the same intrinsic
geometry $\sigma_{ab}$, whether induced by $h_{ij}$ or by
$\barh_{ij}$. We show this in the next subsection. Notice that 
under the transformation $(r,\theta,\phi) \rightarrow x^k$
in Eq.~(\ref{firstprime}), one does not recover Eq.~(\ref{BOM}), 
of course. Rather one obtains
\begin{equation}
h_{ij}(x^k) = \barh_{ij}(x^k) + \Delta h_{ij}(x^k)\, ,
\label{notBOM}
\end{equation}
where $\barh_{ij}(x^k) \neq f_{ij}$ in general.
Eqs.~(\ref{firstprime}) and (\ref{notBOM}) are in fact the same
unique splitting with the flat background $\barSigma$ defined by 
the $B$ embedding, only the coordinates differ. Eq.~(\ref{BOM}) is 
a metric splitting with respect to the flat space $E^3$ defined by 
asymptotically Cartesian coordinates, while Eq.~(\ref{notBOM}) 
represents a different decomposition into background and 
perturbation parts, one defined with respect to the different
Euclidean space $\barSigma$. Later we will have need to consider the 
fall--off for $\Delta h_{ij}(x^k)$  in (\ref{notBOM}) as
$r = (x^k x_k)^{1/2} \rightarrow \infty$ which is also defined 
in terms of $\delta h_{ij}$ through the as yet unknown 
expansion
\begin{equation}
\barh_{ij}(x^k) = f_{ij} + \delta \barh_{ij}(x^k)\, , 
\label{perturb-barh}
\end{equation}
with respect to $f_{ij}$ in $x^k$ coordinates. Now, it is not 
evident that the fall--off for $\delta \barh_{ij}(x^k)$ coincides 
with $\delta h_{ij}(x^k)$ in (\ref{BOM}). However, as shown
below, these fall-offs are qualitatively the same.

\subsection{Isometric embedding of a slightly deformed two--sphere 
into a flat space}

In this subsection we solve the problem of removing a distant 
large--$r$ two--sphere from an asymptotically flat slice $\Sigma$ 
and isometrically embedding it into a flat space $\barSigma$. 
Our method of solution is a perturbative one. Solution of this 
problem defines the transformation between asymptotically Cartesian 
coordinates $x^k$ belonging to $\Sigma$ and Cartesian coordinates
$X^k$ belonging to the flat space $\barSigma$ of the embedding.

Dropping the lowest order term, we rewrite 
the line--element (\ref{BOM}) associated with the 
general $\Sigma$ metric $h_{ij}$ in terms of the polar coordinates
$(r\, ,\theta\, ,\phi)$ associated with $x^k$. Then, fixing
$r = r_0$, with $r_0 \gg 1$ some large constant, we define a large 
two--surface $B(r_0)$, which we refer to as a {\em slightly 
deformed two--sphere}. From the $\Sigma$ line--element rewritten in 
polar coordinates, we may obtain the line element for our slightly
deformed two--sphere,
\begin{eqnarray}
             ds^2_B &\sim&   r^2_0
                             \left(1 
                           + \frac{2\alpha(\theta,\phi)}{r_0} 
                           + \frac{2{\cal  A} 
                             (\theta,\phi)}{r_0^{1+\varepsilon}}
                             \right)d\theta^2
                           + r_0^{2} \left(
                             \frac{2\gamma(\theta,\phi)}{r_0}
                           + \frac{2{\cal  G} 
                             (\theta,\phi)}{r_0^{1+\varepsilon}}
                             \right)d\theta
                             d\phi
                             \nonumber \\
                    &    & +\, r_0^{2} \sin^2\theta
                             \left(1 
                           + \frac{2\beta(\theta,\phi)}{r_0}
                           + \frac{2{\cal  B} 
                             (\theta,\phi)}{r_0^{1+\varepsilon}}
                             \right)
                             d\phi^2\, .
\label{BOM-B}
\end{eqnarray}
It is easy to conclude that under the parity transformation
${\cal P}(x^k) = -x^k$ the spherical coordinates behave as 
${\cal P}(r\,,\theta\, ,\phi) = (r\, ,\pi -\theta\, ,\phi+\pi)$.
Then, keeping in mind that the $a_{ij}$ in (\ref{BOM}) are of even 
parity and the coefficients in (\ref{BOM-B}) are held as 
independent, one sees that $\alpha$ and $\beta$ are of the even
parity, whereas $\gamma$ is of odd parity.

Next, we write down the line--element for the Euclidean space
$\barSigma$ in the corresponding spherical polar coordinates,
\begin{equation}
 ds^2_{\barSigma} = dR^2
                  + R^2\left(d\Theta^2 
                  + \sin^2 \Theta d\Phi^2\right)\, .
\label{Euclidean}
\end{equation}
Our goal is to isometrically embed $B(r_0)$ defined by 
(\ref{BOM-B}) into the flat space $\barSigma$ with line--element
(\ref{Euclidean}). We posit the existence of a coordinate 
transformation with asymptotic form
\begin{subequations}
\label{RThetaPhi}
\begin{eqnarray}
R/r & \sim & 1 + \frac{f(\theta,\phi)}{r}+ \frac{F
(\theta,\phi)}{r^{1+\varepsilon}}  \label{RThetaPhia} \\
\Theta & \sim &\theta + \frac{g(\theta,\phi)}{r}+ \frac{G
(\theta,\phi)}{r^{1+\varepsilon}}  \label{RThetaPhib} \\
\Phi & \sim & \phi + \frac{h(\theta,\phi)}{r}+\frac{H
(\theta,\phi)}{r^{1+\varepsilon}}\, , \label{RThetaPhic}
\end{eqnarray}
\end{subequations}
which features as yet undefined angular functions. Substitution of
the transformation (\ref{RThetaPhi}) into (\ref{Euclidean}) 
yields another expression for the flat metric consistent with 
\bom fall-off. We again fix $r = r_0$ in the resulting expression
to define a two--surface $B(r_0)$ and demand that its corresponding
line--element
\begin{eqnarray}
        ds^2_{B} & \sim  &  r^2_0\left[1
                            +\frac{2}{r_0}
                             \left(f 
                            + \frac{\partial g}{\partial\theta} 
                             \right)
                            +\frac{2}{r_0^{1+\varepsilon}} 
                             \left(F 
                            +\frac{\partial G}{\partial\theta}
                             \right)
                             \right]d\theta^2 
                             \\
                       & & +\, r^2_0
                             \left[\frac{2}{r_0}
                             \left(
                             \frac{\partial g}{\partial\phi}
                            +\sin^2\theta~\frac{\partial h}{\partial\theta}
                             \right)
                            +\frac{2}{r_0^{1+\epsilon}}
                             \left(\frac{\partial G}{\partial\phi}   
                            +\sin^2\theta~\frac{\partial H}{\partial\theta}
                             \right)
                             \right] d\theta d\phi
                             \nonumber \\
                        & & +\, r^2_0\sin^2\theta 
                             \left[1
                            +\frac{2}{r_0}
                             \left(f
                            +g\cot\theta 
                            +\frac{\partial h}{\partial\phi} 
                             \right)
                            +\frac{2}{r_0^{1+\varepsilon}}
                             \left(F+G\cot\theta
                            +\frac{\partial H}{\partial\phi}
                             \right)\right]d\phi^2
                             \nonumber
\end{eqnarray}
matches the line--element (\ref{BOM-B}) just given. Such matching 
is tantamount to solving the isometric embedding problem.
As $1/r_0$ is here a small parameter, we shall --after peeling off an
overall $r_0^2$ factor-- explicitly consider the
$r_0^{-1}$ and  $r_0^{-1-\epsilon}$ orders and implicitly consider
the $r^{-2}$ order. Hence our solution of the isometric embedding problem
will only be approximate. 

Balancing terms at order $r_0^{-1}$, we obtain the system
\begin{subequations}\label{system-r0}
\begin{eqnarray}
  f + \frac{\partial g}{\partial\theta}              & = & \alpha
                                              \label{system-r0a}\\
  f + g \cot \theta 
    + \frac{\partial h}{\partial\phi}                & = & \beta
                                               \label{system-r0b}\\
      \frac{\partial g}{\partial\phi} 
    + \sin^2\theta~\frac{\partial h}{\partial\theta} & = & \gamma\,  ,
\label{system-r0c}
\end{eqnarray}
\end{subequations}
while for the case $\varepsilon <1$ a similar balance of terms at 
order $r_0^{-1-\varepsilon}$ 
yields the system
\begin{subequations}\label{system-r0e}
\begin{eqnarray}
  F + \frac{\partial G}{\partial\theta}              & = & {\cal A}
                                               \label{system-r0ea}\\
  F + G \cot\theta + \frac{\partial H}{\partial\phi} & = & {\cal B}
                                               \label{system-r0eb}\\
      \frac{\partial G}{\partial\phi} 
    + \sin^2\theta~\frac{\partial H}{\partial\theta} & = & {\cal G}\, .
\label{system-r0ec}
\end{eqnarray}
\end{subequations}
In the case $\varepsilon = 1$ the system is at order $r_0^{-2}$ and 
the {\sc rhs} of each equation in
(\ref{system-r0e}) is modified by addition of known functions of 
$f$, $g$, $h$ and their derivatives: ${\cal A} \mapsto {\cal A} 
+ {\cal A}'(f,g,h)$, ${\cal B} \mapsto {\cal B} + {\cal B}'(f,g,h)$,
${\cal G} \mapsto {\cal G} + {\cal G}'(f,g,h)$. Of course, we would
assume that $f$, $g$, $h$ were obtained via resolution of the first 
system (\ref{system-r0}) before turning to solve the modified second 
system. In the case $\varepsilon < 1$ we may also obtain a system at 
order $r_0^{-2}$, yet a third system and one of the same type. This 
is possible because we have included the extra terms $b_{ij}$ and
$c_{ij}$ in Eq.~(\ref{BOM}) and therefore use a more
detailed expansion than the one \bom use. Considering all of these
systems together, including the possible third unwritten system 
at  order $r_0^{-2}$, we may solve the embedding equations 
asymptotically to a high enough order to ensure that
$\sigma_{ab} - \barsigma_{ab} = O(r^{-\varepsilon})$, in turn 
giving the following crucial result: $\Delta\sigma^{ij} = 
O(r^{-2-\varepsilon})$. 

We may consider our isometric embedding problem solved (to the
required accuracy) if there is a combined solution to the systems
(\ref{system-r0}) and (\ref{system-r0e}) as well as the third 
system we just mentioned. In {\em Appendix A.2} we examine the 
system (\ref{system-r0}), finding a solution assuming simple
compatibility conditions for the coefficients
in (\ref{BOM-B}). The remaining systems are formally similar 
and may be examined analogously. Note that conditions of regularity
on the coefficients $\alpha$, $\beta$, and $\gamma$ ({\em intrinsic}
$B$ geometry) are expected and should mirror conditions on the
full metric $\sigma_{ab}$ which are necessary and sufficient for 
existence of a (suitably unique) solution to the full embedding 
equations. If the Ricci scalar (twice the Gaussian 
curvature) of $B$ is everywhere positive (as is the case here), 
the task of isometrically embedding $B$ in Euclidean space is 
{\em Weyl's problem}, a classic embedding problem
of differential geometry in the large. In the
most robust formulation of the problem known to us,
that due to Heinz \cite{Heinz}, existence
of such an embedding is guaranteed if the $B$ metric
coefficients are twice continuously differentiable.
Uniqueness of the embedding (up to Euclidean motions)
follows from the Cohn--Vosson theorem. \cite{Spivak}

First performing the trivial transformation from the system 
$(R,\Theta,\Phi)$ to the corresponding Cartesian coordinates $X^k$,
we now seek, subject to the assumption of \bom fall--off, 
the behavior of the transformation between $X^k$ and $x^k$. In fact,
the transformation (\ref{RThetaPhi}) can be rewritten as [see 
(\ref{XYZ}) in {\em Appendix A.2} and the text thereafter] as
\begin{equation}
X^k \sim x^k +\, {}^{0}\xi^k(\nu^l) 
             +\, {}^{\varepsilon}\xi^k r^{-\varepsilon}
             +\, {}^{1}\xi^k r^{-1}\, ,
\label{expansion-X}
\end{equation}
whence
\begin{equation}
\partial X^k/\partial x^i     \sim   \delta^k{}_i 
                                 + {\cal A}^k{}_i(\nu^l)r^{-1}
                                 + {\cal B}^k{}_i
                                   r^{-1-\varepsilon}
                                 + {\cal C}^k{}_i r^{-2}\, .
\label{expansion-X/x}
\end{equation}
The angular functions ${}^{0}\xi^k(\nu^l)$ and ${\cal A}^k{}_i(\nu^l)$ 
are of odd and even parity respectively, while the angular functions 
${}^{\varepsilon}\xi^k$, ${}^{1}\xi^k$, ${\cal B}^k{}_i$, and 
${\cal C}^k{}_i$ are not of definite parity. Note that the 
${\cal A}^k{}_i$ and ${\cal B}^k{}_i$ here are not the $\cal A$ and 
$\cal B$ appearing in the system (\ref{system-r0e}).

\subsection{Asymptotic expansions for $\Delta h_{ij}$, $\Delta 
q^{ij}$, and $\Delta n^i$}
With respect to the \bom metric splitting (\ref{BOM}), we first
collect the following standard and easily derived formulae:
\begin{subequations}\label{hqMnexpansions}
\begin{eqnarray}
h_{ij} = & f_{ij} + \delta h_{ij} & 
             \sim f_{ij} 
            + a_{ij}r^{-1} 
            + b_{ij}r^{-1-\varepsilon}
            + c_{ij}r^{-2}
\label{hqMnexpansionsa}\\
h^{ij} = & f^{ij} + \delta q^{ij} & 
              \sim f^{ij} 
            - a^{ij}r^{-1}
            - b^{ij}r^{-1-\varepsilon}
            - (c^{ij}
            - a^{i}{}_l a^{lj})r^{-2}
              \label{hqMnexpansionsb}  \\
M      = & 1 + \delta M           &
              \sim 1 + \half a_{\nu\nu}r^{-1} 
               + \half b_{\nu\nu}r^{-1-\varepsilon}
               + \half (c_{\nu\nu} - a_\nu{}^i a_{i\nu} 
               + {\textstyle \frac{3}{4}} a_{\nu\nu}^2)r^{-2}
              \label{hqMnexpansionsc} \\
n^i    = & \nu^i + \delta n^i     & 
             \sim \nu^i 
            + \half
              (\nu^i a_{\nu\nu}   
            - 2a^{i}{}_{\nu})r^{-1}
            + \half (\nu^i b_{\nu\nu}
            - 2b^{i}{}_{\nu})r^{-1-\varepsilon}
              \nonumber \\
       &   & + \half
             \left[
             \nu^i(c_{\nu\nu} - a_\nu{}^k a_{k\nu}
               + {\textstyle \frac{3}{4}} a_{\nu\nu}^2)
               - 2c^i{}_\nu + 2 a^{ik}a_{k\nu}
               - a^i{}_\nu a_{\nu\nu}
             \right]r^{-2}
              \label{hqMnexpansionsd}
\end{eqnarray}
\end{subequations}
with the equalities indicating that $\delta h_{ij}$, $\delta q^{ij}$, 
$\delta M$, and $\delta n^i$ contain {\em all} orders of 
perturbations. Moreover, on the rightmost side of the equations all 
indices have been raised or lowered with the trivial flat $f_{ij}$ 
metric and a subscript $\nu$ indicates contraction with $\nu^i$ (for 
example $a^i{}_{\nu} = a^{ik}\nu_k$). 

With $\barh_{ij}(X^k) = f_{ij}$ and the expansion (\ref{expansion-X/x}),
we find the sought--for expansion (\ref{perturb-barh}) of
{\sc b}\'o{\sc m}--type for the $\barh_{ij}$ background metric,
\begin{equation}
\barh_{ij}  \sim f_{ij}
            + {\sf a}_{ij}(\nu^k)r^{-1}
            + {\sf b}_{ij}r^{-1-\varepsilon}
            + {\sf c}_{ij}r^{-2}\, ,
\label{barhbomexp1}
\end{equation}
where now all subdominant terms are pure--gauge and given by
\begin{subequations}\label{barhbomexp2}
\begin{eqnarray}
\bara_{ij} & = & {\cal A}_{ij} + {\cal A}_{ji}
\label{barhbomexp2a}\\
\barb_{ij} & = & {\cal B}_{ij} + {\cal B}_{ji}
\label{barhbomexp2ab} \\
\barc_{ij} & = & {\cal A}_{ki} {\cal A}^k{}_j +
                   {\cal C}_{ij} + {\cal C}_{ji}\, .
\label{barhbomexp2c}
\end{eqnarray}
\end{subequations}
Note that $\bara_{ij}$ has even parity, while both $\barb_{ij}$
and $\barc_{ij}$ are not of definite parity.
From the expansion (\ref{barhbomexp1}) we obtain identities for 
$\barh_{ij}$, $\barh^{ij}$,
$\barM$, and $\barn^i$ which are token identical to
those given in Eqs.~(\ref{hqMnexpansions}). Moreover, these
mirror identities also have the same parity behavior as those
found in Eqs.~(\ref{hqMnexpansions}). Substitution of 
Eqs.~(\ref{BOM}) and (\ref{barhbomexp1}) into Eq.~(\ref{notBOM}) 
and these considerations show that
\begin{subequations}
\begin{eqnarray}
\Delta h_{ij} & = & \delta h_{ij} - \delta\barh_{ij} 
\\
\Delta q^{ij} & = & \delta q^{ij} - \delta\barq^{ij} 
\\
\Delta n^{i} & = & \delta n^i - \delta\barn^{i} 
\end{eqnarray}
\end{subequations}
each possess precisely the same fall--off and parity properties 
as $\delta h_{ij}$, $\delta q^{ij}$, and $\delta n^i$, although
of course they do differ term--by--term from these.

\subsection{Energy integral}
We turn now to the detailed comparison between the Brown--York and  
\bom integrals. If we naively insert the identities (\ref{identity})
in terms of $\Delta$ perturbations into the Brown--York integral 
$H_B$, then we arrive at expressions appearing to be the {\sc
adm} and \bom integrals. However, one cannot make such a direct 
comparison, as we have expressed $H_B$ in terms of $\Delta$ 
perturbations with respect to reference Euclidean space $\barSigma$, 
whereas the \bom integral is expressed in terms of $\delta$ perturbations
with respect to the Euclidean space $E^3$ defined by the asymptotically
Cartesian coordinates. 
An elegant way to achieve the comparison would appeal to the
technique of gauge (inner) transformations \cite{PP88} based on the
theory of continuous transformations \cite{Eisenhart}.
However, we adopt a more cumbersome but straightforward approach.

Let us first consider energy scenario. As demonstrated later in the
text around Eq.~(\ref{detsigma}) below, 
the 
proper area--element for a level--$r$ two--surface 
$B$ obeys
\begin{equation}
d^2x\sqrt{\sigma} = d\Omega r^2\big[1+ \half(a^i{}_i -
a_{\nu\nu})r^{-1}
+ O(r^{-1-\varepsilon})\big]\, ,
\label{elementofB}
\end{equation}
where again $d\Omega$ denotes the area--element of the unit sphere. 
Only the leading behavior of this equation is relevant for
the energy scenario at hand; however, the next--to--leading order is 
relevant for the center--of--mass scenario, since a
lapse $N$ growing like $r$ will ``sample'' this next--to--leading order.
Hence we keep it here for future use. Our results from the last subsection
further imply that
\begin{equation}
\sqrt{\sigma} - \sqrt{\barsigma} = O(r^{-\varepsilon})\, .
\end{equation}
Therefore, as far as all energy scenario limits and even 
center--of--mass scenario limits are concerned, we may view 
$d^2x\sqrt{\sigma}$ as the area--element induced either by $h_{ij}$ or 
$\barh_{ij}$. 

Since the area--element grows like $r^2$ and for now
$N \sim 1$, we see that only leading $O(r^{-2})$ terms in the expression
(\ref{identity1}) for $k-\bark$ will contribute to the $r\rightarrow\infty$ 
limit of $H_B$. It is readily seen that the last four terms in 
(\ref{identity1}) can not contribute to the limit. Indeed, symbolically 
$\barD = \partial + \barGamma$, where $\barGamma$ Christoffel terms are
$O(r^{-2})$, a fact stemming from the derived fall--off (\ref{barhbomexp1})
for $\barh_{ij}$.
Therefore, $\barD_k$ differentiation --like $\partial_k$ differentiation-- 
drops fall--off by one power of inverse radius (and reverses parity). This 
shows that the third, fourth and fifth terms are $O(r^{-3})$. Finally, our 
perturbative solution to the embedding equations has shown the sixth and 
final term to be $O(r^{-3-\varepsilon})$.

Now considering the remaining two factors (the first two) on the 
{\sc rhs} of (\ref{identity1}), we drop some more subdominant terms and 
write
\begin{equation}
k-\bark \sim \half \barn^i h^{kl}
\big(\partial_k\Delta h_{il} - \partial_i \Delta h_{kl}\big)
+ \nu_k\partial_l\big(\barn^{[l}\Delta n^{k]}\big)\, .
\end{equation}
The product of the last factor and the leading contribution to
the area--element (\ref{elementofB}) integrates to zero via Stokes'
theorem. Whence for the energy scenario we have that 
\begin{equation}
H_{B} \sim \frac{1}{16\pi}\int_{B}d^2x\sqrt{\sigma}\barn^i h^{kl}
\big(\partial_k\Delta h_{il} - \partial_i \Delta h_{kl}\big)\, ,
\end{equation}
a result which yields
\begin{equation}
H_{B} \sim \frac{1}{16\pi}\int_{B}d^2x\sqrt{\sigma}n^i h^{kl}
\big(\partial_k h_{il} - \partial_i h_{kl}\big)-
\frac{1}{16\pi}\int_{B}d^2x\sqrt{\sigma}\barn^i \barh^{kl}
\big(\partial_k \barh_{il} - \partial_i \barh_{kl}\big)
\label{HBdiff}
\end{equation}
after some replacements of $n^i$ and $h^{ij}$ by $\barn^i$ 
and $\barh^{ij}$ at the expense of introducing subdominant
terms which are then discarded. Apart from the fact that the 
integration is over the finite surface $B$ and not the sphere
at infinity, the first integral on the {\sc rhs} is the 
{\sc adm} energy belonging to $\Sigma$. Likewise, in the limit
the second integral is the {\sc adm} energy belonging to the 
vacuum slice $\barSigma \simeq E^3$. To see this, merely replace 
$\sqrt{\sigma}$ with $\sqrt{\barsigma}$ at the expense of 
introducing subdominant terms. Because 
we work with physically reasonable \bom or \rt fall--off, we 
expect that this second integral then vanishes in the limit. That it 
indeed does may also be verified via direct calculation.
With Eqs.~(\ref{expansion-X/x},\ref{barhbomexp1},\ref{barhbomexp2})
we get
\begin{equation}
\barh_{ij} \sim f_{ij} + \partial_i \, ^{0}\xi_j + \partial_j\, 
^{0}\xi_i\, .
\end{equation}
Whence to leading order the integrand of the second integral in 
Eq.~(\ref{HBdiff}) is 
$\nu^i\partial^l \big(\partial_{[l}\,{}^{0}\xi_{i]}\big)$, and again 
the leading order of the product of this term with the area--element 
integrates to zero via Stokes' theorem.

\subsection{Center--of--mass integral}

Evaluation of the integral (\ref{BYintegral}) for the scenario with
$N$ behaving as in Eq.~(\ref{N}) is quite a bit more subtle. The 
integral is naively divergent in the $r \rightarrow \infty$ limit, so 
we must keep track of both divergent terms (which only vanishes upon 
invocation of the Hamiltonian constraint) and finite terms. Let us 
turn now to the identity (\ref{identity2}) for $N(k-\bark)$ and first 
dispatch the terms which most obviously do not contribute to the 
limit. First, as the last two terms are 
$O(r^{-2-\varepsilon})$ they do not contribute. The fourth, fifth,
sixth, and seventh terms are each $O(r^{-2})$. Nevertheless, none of 
these terms contribute to the limit, since for each parity conditions 
ensure that all leading terms integrate to zero. For example, consider 
the fourth term, $\half N\barn^i \Delta h_{il} \barD_k\Delta q^{kl}$.
As argued just above, $\barD_k$ differentiation drops fall--off and
reverses parity, so $\barD_k\Delta q^{kl}$ is $O(r^{-2})$ and leading 
odd--parity. The product $\barn^i \Delta h_{il}$ is clearly
$O(r^{-1})$ and also leading odd--parity. Since the lapse here is $O(r)$
and leading odd--parity, we see that the full term is leading
odd--parity, whence the product of this term with the leading contribution
to the area--element integrates to zero.

Focus now on the third term, $\barn_k\barD_l(N\barn^{[l}\Delta n^{k]})$,
on the {\sc rhs} of (\ref{identity2}), writing it as
\begin{equation}
\barn_k\barD_l(N\barn^{[l}\Delta n^{k]}) = 
\barn_k\partial_l(N\barn^{[l}\Delta n^{k]}) 
+ \barn_k(N\barn^{[p}\Delta n^{k]})\barGamma^l{}_{pl}\, .
\end{equation}
The term involving the Christoffel symbol is $O(r^{-2})$ and 
with leading odd parity, essentially because the $\barGamma^k{}_{pl}$ 
are themselves $O(r^{-2})$ and of leading odd parity. Therefore, this
term does not contribute to the limit. The first term 
on the {\sc rhs} also makes no contribution. To show this, first
recall the discussion above and introduce the odd--parity term
\begin{equation}
A^i =         \half(\nu^i a_{\nu\nu}
            - 2a^{i}{}_{\nu}- \nu^i \bara_{\nu\nu}
            + 2\bara^{i}{}_{\nu})\, ,
\label{sancapitalA}
\end{equation}
so that $\Delta n^i \sim A^i r^{-1}$. Next, expand 
$N\barn^{[l}\Delta n^{k]}$, the factor within the parenthesis, as 
follows
\begin{equation}
N\barn^{[l}\Delta n^{k]} = \beta^{\perp}{}_{\nu}\,\nu^{[l}A^{k]}
                         + \lambda^{[lk]} + \rho^{[lk]} 
                         + O(r^{-1-\varepsilon}) \, ,
\end{equation}
where $\beta^{\perp}{}_{\nu} = \beta^{\perp}{}_j\nu^j$ is $O(r^0)$ and
$\lambda^{[lk]}$ and $\rho^{[lk]}$ are 
respectively $O(r^{-\varepsilon})$ and $O(r^{-1})$. From the last equation, 
infer that
\begin{equation}
\barn_k \partial_l(N\barn^{[l}\Delta n^{k]})
= \barn_k \partial_l(\beta^{\perp}{}_{\nu}\,\nu^{[l}A^{k]})
+ \nu_k\partial_l(\lambda^{[lk]} + \rho^{[lk]}) 
+ O(r^{-2-\varepsilon})\, .
\end{equation}
With the analog of (\ref{hqMnexpansionsc}) for $\barM$ in
$\barn_k = \barM\nu_k$, the fact that $\beta^\bot{}_\nu$ is $O(1)$ and
of odd parity, and Eq.~(\ref{sancapitalA}), we conclude that on
the {\sc rhs} the first and only worrisome term includes $O(r^{-1})$ 
and $O(r^{-2})$ pieces (both potentially contributing to the integral) 
which are of odd parity and thus integrate to zero.

We drop those terms in Eq.~(\ref{identity2}) shown to make
no contribution to the limit, thereby reaching
\begin{equation}
H_B \sim \frac{1}{16\pi}\int_{B}d^2x\sqrt{\sigma}
\left[N \barn^i h^{kl}\left(\partial_k
                               \Delta h_{il}
                             - \partial_i \Delta h_{kl}\right)
                             - \left(\barn^k h^{il}\Delta h_{ik}
                             - \barn^l h^{ik}\Delta h_{ik}\right)
                               \partial_{l} N\right]\, . 
\end{equation}
To get this last equation we have also replaced $\barD_i$ 
differentiation with $\partial_i$ differentiation, a step easily seen 
as permissible via fall--off and parity arguments. 
We now write $\Delta h_{ij} = 
\delta h_{ij} -\delta \barh_{ij}$, in order to cast the last equation 
into the form
\begin{eqnarray}
H_B & \sim & \frac{1}{16\pi}\int_{B}d^2x\sqrt{\sigma}
             \left[N n^i h^{kl}\left(\partial_k h_{il}
                             - \partial_i h_{kl}\right)
                             - \left(n^k h^{il}\delta h_{ik} 
                             - n^l h^{ik}\delta h_{ik}\right)
                               \partial_{l} N\right]
\nonumber \\
    &  -  &  \frac{1}{16\pi}\int_{B}d^2x\sqrt{\sigma}
             \left[N \barn^i \barh^{kl}\left(\partial_k
                               \barh_{il}
                             - \partial_i\barh_{kl}\right)
                             - \left(\barn^k \barh^{il}\delta\barh_{ik} 
                             - \barn^l \barh^{ik}\delta \barh_{ik}\right)
                               \partial_{l} N\right]\, .
\label{diffofBOMs}
\end{eqnarray}
To reach this equation, we have made several swaps of $\barh^{ij}$ for
$h^{ij}$ and $n^i$ for $\barn^i$. Such swaps are permissible, since
they result in the introduction of $O(r^{-2})$ terms in the integrand
which are then seen to integrate to zero via parity arguments. One 
recognizes the first integral as the \bom integral 
$M^{\perp}_B(N)$ belonging to $\Sigma$. It is easy to infer that the 
second integral on the {\sc rhs} of Eq.~(\ref{diffofBOMs}) must vanish
in the $r\rightarrow\infty$ limit. 
First replace $\sqrt{\sigma}$ with $\sqrt{\barsigma}$, a step which 
does not affect the limit. The resulting integral $\barM^{\perp}_B(N)$ 
is the \bom integral for $\barh_{ij}$ and the vacuum slice 
$\barSigma\simeq E^3$; hence we also expect this integral to vanish
in the limit as $\barh_{ij}$ obeys physically reasonable fall--off.

Let us quickly establish the vanishing of \bom integral 
when evaluated on the flat metric $\barh_{ij}$ belonging to the 
vacuum slice $\barSigma$. Now, the full integrand, the sum of four 
terms, is $O(r^{-1})$ and of leading odd parity. 
Consider the product of its 
leading term with the area--element (\ref{elementofB}).
It is only this leading term of the integrand which ``sees'' the
next--to--leading term in the area element; that is to say, the
product of the leading term of the integrand with the
next--to--leading term of the area element potentially contributes
to the integral. But we see that this potential contribution
integrates to zero via parity arguments, whence in the calculations
to follow we may work solely with the leading round--sphere term
$d\Omega r^2$ of the area element.

Without affecting the $r\rightarrow\infty$ limit, the integrand for 
the \bom integral of $\barh_{ij}$ may be replaced by the sum of the 
following four terms:
\begin{subequations}\label{fourterms}
\begin{eqnarray}
{\rm term1} & = & (\beta^\perp{}_l x^l)\nu^k f^{ij}\partial_j \barh_{ik}\\
{\rm term2} & = & - (\beta^\perp{}_l x^l) 
                  \nu^k f^{ij}\partial_k \barh_{ij}\\
{\rm term3} & = & - \beta^\perp{}_l \nu^i f^{lj} \delta \barh_{ij}\\
{\rm term4} & = & \beta^\perp{}_l \nu^l f^{ij} \delta \barh_{ij} \, .
\end{eqnarray}
\end{subequations}
To see why this is the case, consider, for example, the term 
\begin{eqnarray}
N \barn^k \barh^{ij}\partial_j \barh_{ik} 
            & = & N \nu^k f^{ij}\partial_j \barh_{ik} 
              +   N \delta \barn^k f^{ij}\partial_j \barh_{ik}
              \nonumber  \\
            & + &  
                  N \nu^k \delta \barq^{ij}\partial_j \barh_{ik}  
              +   N \delta \barn^k \delta \barq^{ij}\partial_j 
                  \barh_{ik} \, .
\end{eqnarray}
On the {\sc rhs} the last term is $O(r^{-3})$ while the second and third
terms are each $O(r^{-2})$ and off odd parity, whence none of these terms
contribute to the limit. Finally, the first term on the {\sc rhs} is
\begin{equation}
N \nu^k f^{ij}\partial_j \barh_{ik} = 
(\beta^\perp{}_l x^l)\nu^k f^{ij}\partial_j \barh_{ik} + 
\alpha^{-} \nu^k f^{ij}\partial_j \barh_{ik} + 
O(r^{-2-\varepsilon})\, ,
\end{equation}
where the middle term on the {\sc rhs} is $O(r^{-2})$ and of odd 
parity, so on the {\sc rhs} the sole limit--contributing term is the 
first one.

Eqs.~(\ref{expansion-X/x},\ref{barhbomexp1},\ref{barhbomexp2c})
allow us to write
\begin{eqnarray}
\barh_{ij} & = & 
f_{ij} + \pounds_{\xi}f_{ij} + ({\rm E.P.T.})r^{-2} + 
O(r^{-2-\varepsilon})
\nonumber \\
& = & 
f_{ij} + \partial_i\xi_j +\partial_j\xi_i + 
({\rm E.P.T.})r^{-2} + O(r^{-2-\varepsilon})\, .
\end{eqnarray}
Again, (E.P.T.) stands for generic terms of even parity, and the
one here stems from the ${\cal A}_{ki}{\cal A}^k{}_j$ in 
Eq.~(\ref{barhbomexp2c}). 
When coupled with simple parity and fall--off arguments, 
this expression for $\barh_{ij}$ shows that we may replace $\delta 
\barh_{ij}$ terms in Eqs.~(\ref{fourterms}) with 
$\pounds_\xi f_{ij}$. Therefore, up to terms not contributing in
the $r\rightarrow\infty$ limit,
\begin{eqnarray}
\lefteqn{
{\rm term1} + {\rm term2} + {\rm term3} + {\rm term4}}
& & \nonumber \\
& = & 2(\beta^\perp{}_l x^l)\nu^k f^{ij}\partial_i\partial_{[j} \xi_{k]}
- \beta^\perp{}_l \nu^i f^{lj}
(\partial_i \xi_j + \partial_j \xi_i)
+ 2\beta^\perp{}_l \nu^l f^{ij}\partial_i \xi_j
\nonumber \\
& = & 
2\nu^k\partial_i
\big[(\beta^\perp{}_l x^l) f^{ij}\partial_{[j} \xi_{k]}\big]
- 2\beta^\perp{}_l \nu^i f^{lj}\partial_j \xi_i
+ 2\beta^\perp{}_l \nu^l f^{ij}\partial_i \xi_j
\nonumber \\
& = &
2\nu_k\partial_i
\big[(\beta^\perp{}_l x^l)\partial^{[i} \xi^{k]}
- 2\beta^{\perp [i} \xi^{k]}\big]\, ,
\end{eqnarray}
whence ---to the relevant order--- the full integrand
integrates to zero via Stokes' theorem.

\section{$H_B$ and the curvature integral $M_\Re$}
In this section we again evaluate the main surface integral 
(\ref{BYintegral})
in the large--sphere limit towards {\sc spi}, this time 
showing that
\begin{equation}
H_{B} \sim
       \frac{1}{16\pi}\sqrt{\frac{A}{4\pi}}
       \int_{B}d^{2}x
       \sqrt{\sigma} N\sigma^{\mu\nu}\sigma^{\lambda\kappa}
       \Re_{\mu\lambda\nu\kappa}
\label{mainlimit}
\end{equation}
as $r \rightarrow \infty$. We have discussed the integral
on the {\sc rhs} in the paragraph after Eq.~(\ref{mainlimit0}) 
where it is denoted $M_\Re$.
This result holds for both the energy scenario with $N \sim 1$ and 
the center--of--mass scenario with $N$ given by Eq.~(\ref{N}) 
discussed in {\em Section II}.

We will verify (\ref{mainlimit}) by establishing the following
results. First,
\begin{equation}
      k - k|^{\rm ref} \sim
                           \,{}^3\!
                           \left[\half\sigma^{\mu\nu}
                           \sigma^{\lambda\kappa}
                           \Re_{\mu\lambda\nu\kappa}\right]r^{-2}\, ,
\label{firstresult}
\end{equation}
where the superscript 3 means take the coefficient of the leading 
$O(r^{-3})$ term in a radial expansion of the curvature component 
$\half \sigma^{\mu\nu}\sigma^{\lambda\kappa}\Re_{\mu\lambda\nu\kappa}$. 
The coefficient ${}^3\!\left[\half\sigma^{\mu\nu}\sigma^{\lambda\kappa}
\Re_{\mu\lambda\nu\kappa}\right]$ happens to be of even parity.
Second, Eq.~(\ref{elementofB}) can be written
\begin{equation}
d^2 x\sqrt{\sigma} = r^2 d\Omega\left[1 + ({\rm E.P.T.})r^{-1} 
+ O(r^{-1-\varepsilon})\right]\, ,
\label{secondresult}
\end{equation}
where we use E.P.T. to stand for generic terms of even parity. 
Important in itself, this result also shows that
$\sqrt{A/4\pi} \sim r + O(r^{-1})$. 
Third,
\begin{equation}
       k - k|^{\rm ref} \sim 
                             \big({\rm E.P.T.}\big)r^{-2} 
                           + {}^{(3+\varepsilon)}\!
                             \left[\half
                             \sigma^{\mu\nu}
                             \sigma^{\lambda\kappa}
                             \Re_{\mu\lambda\nu\kappa}\right] 
                             r^{-2-\varepsilon}
                           + \big({\rm E.P.T.} 
                           + {}^4\!\left[\half
                             \sigma^{\mu\nu}
                             \sigma^{\lambda\kappa}
                             \Re_{\mu\lambda\nu\kappa}
                             \right]\big)r^{-3}\, .
\label{thirdresult}
\end{equation}
The E.P.T. coefficient sitting before the $r^{-2}$ is again
${}^3\!\left[\half\sigma^{\mu\nu}\sigma^{\lambda\kappa}
\Re_{\mu\lambda\nu\kappa}\right]$, and the term 
${}^{(3+\varepsilon)}\!\left[\half\sigma^{\mu\nu}
\sigma^{\lambda\kappa} \Re_{\mu\lambda\nu\kappa}\right]$
is not of definite parity. When coupled with by now standard 
arguments, these three results establish Eq.~(\ref{mainlimit}) 
subject to the assumptions of either the energy scenario or 
center--of--mass scenario.

Before turning to detailed calculations, let us lay some groundwork and fix
more notations. We may write the $\Sigma$ metric in an {\sc adm}--like
form
\begin{equation}
h_{ij} {\rm d}x^i {\rm d}x^j = 
M^2 {\rm d}r^2 + \sigma_{ab}({\rm d}x^a + W^a {\rm d}r)
                            ({\rm d}x^b + W^b {\rm d}r)
\, ,
\label{Sigmametric}
\end{equation}
where $M$ and $W^a$ are respectively the radial lapse function and
radial shift vector. Take $x^a = (\zeta, \bar{\zeta})$ as coordinates 
on a $B$ surface. Here $\zeta$ ---a single complex coordinate on $B$---
is the stereographic coordinate $e^{{\rm i}\phi}\cot (\theta/2)$ 
belonging to the asymptotically Cartesian coordinates
$x^k$, and $\bar{\zeta}$ is the complex conjugate of $\zeta$. Introduce 
a complex null 
dyad $m^a \partial/\partial x^a$ and $\bar{m}{}^a \partial/\partial x^a$ 
on $B$, chosen so that $m_a \bar{m}{}^a = 1$ and $m_a m^a = 0$. In terms
of the null dyad the $B$ metric is written as
\begin{equation}
\sigma_{ab} = m_a \bar{m}_{b} + \bar{m}_a m_b\, .
\label{Bmetric}
\end{equation}
One advantage of working with a complex null $B$ dyad is that results
derived for one frame leg $e_{\hat{3}} = m^{a}\partial/\partial x^a$
yield corresponding results for the other leg $e_{\hat{4}} = \bar{m}^{a} 
\partial/\partial x^a$ under complex conjugation. (See {\em Appendix B.1}
for an explanation as to why we use $\hat{3}$ and $\hat{4}$ as the name
indices here.) Another is that the components $m^{a} m^b k_{ab}$
and $2m^a\bar{m}{}^b k_{ab}$ respectively capture the trace--free and 
trace pieces of the extrinsic curvature tensor $k_{ab}$ associated 
with the radial foliation of $\Sigma$ into nested $B$ two--surfaces.
As an orthonormal co--triad on $\Sigma$ take
\begin{subequations}\label{Sigmacoframe}
\begin{eqnarray}
e^{\vdash} & = & M {\rm d}r 
\\
e^{\hat{3}} = \bar{m}_a ({\rm d} x^a + W^a {\rm d}r) & = & 
\bar{m}_{\zeta} {\rm d}\zeta
+ \bar{m}_{\bar{\zeta}} {\rm d}\bar{\zeta} 
+ W_{\bar{m}} {\rm d}r\, ,
\end{eqnarray}
\end{subequations}
where $W_{\bar{m}} = \bar{m}_{a} W^a$. We may obtain the one--form 
$e^{\hat{4}}$ from $e^{\hat{3}}$ via conjugation. The spatial triad
$e_{\hat{k}}$ dual to and the connection coefficients 
$\omega_{\hat{\jmath}\hat{k}\hat{l}}$ determined by this co--triad are
listed respectively in the appendix Eqs.~(\ref{Sigmaframe}) and
(\ref{Sigmaconnection}). Triad indices $\hat{\jmath},\hat{k},\hat{l},
\cdots$ run over the values $\vdash,\hat{3},\hat{4}$. 

\subsection{Asymptotic expansions}
Recall that $\nu^k = x^k/r$, and with the stereographic coordinate
define the complex direction
\begin{equation}
\mu^i = \sqrt{\textstyle{\frac{1}{2}}}P^{-1}\Big(1-\bar{\zeta}{}^2,
-{\rm i}\big(1+\bar{\zeta}{}^2\big), 2\bar{\zeta}\Big)
\label{muvector}
\end{equation}
discussed in {\em Appendix B.1}. The Cartesian components 
$\mu^i$ obey the relationship
\begin{equation}
f^{ij} = \mu^i \bar{\mu}{}^j + \bar{\mu}{}^i \mu^j + \nu^i \nu^j\, .
\label{numusplitoff}
\end{equation}
With $\mu^i$ and $\nu^i$, we define the following $O(r^0)$ quantities
(and their complex conjugates):
\begin{subequations}\label{aprojection}
\begin{eqnarray}
a_{\nu\nu} & = & \nu^i\nu^j a_{ij}
\\
a_{\nu\mu} & = & \nu^i \mu^j a_{ij}
\\
a_{\mu\bar{\mu}} & = & \mu^i \bar{\mu}{}^j a_{ij}
\\
a_{\mu\mu} & = & \mu^i \mu^j a_{ij}\, .
\end{eqnarray}
\end{subequations}
Both $a_{\nu\nu}$ and $a_{\mu\bar{\mu}}$ are in fact of even parity, 
although this is not obvious for the latter quantity, while $a_{\nu\mu}$ 
and $a_{\mu\mu}$ are not of definite parity. Below we discuss the behavior
of $a_{\nu\mu}$, $a_{\mu\bar{\mu}}$, and $a_{\mu\mu}$ under the parity
operation. 

The expansions for the metric functions $M$ and $W_{m}$ are
\begin{subequations}\label{MandW}
\begin{eqnarray}
M & \sim & 1 + \half a_{\nu\nu} r^{-1} 
\label{MandWa}\\
W_{m} & \sim & 
a_{\nu\mu} r^{-1}
\label{MandWb}\, .
\end{eqnarray}
\end{subequations}
To motivate similar expansions for $m_{\zeta}$ and $m_{\bar{\zeta}}$, 
let us first note that
\begin{equation}
\mu_{i} {\rm d}x^i  =  r\sqrt{2}P^{-1}
{\rm d}\bar{\zeta}\, ,
\end{equation}
which may be demonstrated via the chain rule. Therefore, $\mu_{\zeta} = 
\mu_i \partial x^i/\partial\zeta = 0$ and $\mu_{\bar{\zeta}} = \mu_i 
\partial x^i/\partial\bar{\zeta} = r\sqrt{2}P^{-1}$, although both 
$\partial x^i/\partial\zeta$ and its conjugate are $O(r)$. These results
determine the leading behavior in the expansions for the components of the 
$B$ co--frame,
\begin{subequations}\label{MandM}  
\begin{eqnarray}
m_{\zeta} & \sim & 0\cdot r + \sqrt{\half}P^{-1}a_{\mu\mu}
\\
m_{\bar{\zeta}} & \sim & 
\sqrt{2}P^{-1} r
+ \sqrt{\half}P^{-1}a_{\mu\bar{\mu}} 
\, .
\end{eqnarray}
\end{subequations}
We have carefully tailored the next--to--leading order behavior in the 
expansions (\ref{MandM}) to ensure that
\begin{equation}
m_{b} = \mu_b + 
\half\left(
a_{\mu\mu} \bar{\mu}_{b} + a_{\mu\bar{\mu}} \mu_a
\right)r^{-1}
+ O(r^{-\varepsilon})\, .
\end{equation}
In this equation note that $\mu_a = \mu_i \partial x^i/\partial x^a$ is 
$O(r)$. Now construct $\sigma_{bc}$ via Eq.~(\ref{Bmetric}), thereby
finding
\begin{equation}
\sigma_{bc} = f_{bc} + a_{bc}r^{-1} + O(r^{1-\varepsilon}) \, ,
\label{projectedh}
\end{equation}
where $f_{bc} = \mu_b \bar{\mu}_c + \bar{\mu}_b \mu_c$ is the $O(r^2)$ 
metric on a radius--$r$ round sphere and 
\begin{equation}
a_{bc} = (\mu^i \bar{\mu}_b + \bar{\mu}{}^i \mu_b)
         (\mu^j \bar{\mu}_c + \bar{\mu}{}^j \mu_c) a_{ij}
\end{equation}
is the $O(r^2)$ projection of $a_{ij}$ into the said radius--$r$ sphere.
Note that $\partial x^i/\partial x^b = \mu^i \bar{\mu}_b + 
\bar{\mu}{}^i \mu_b$ as can be shown by Eq.~(\ref{numusplitoff}).
We see therefore that our expansions (\ref{MandM}) for the null co--frame
are consistent, since contraction of Eq.~(\ref{BOMzero1}) 
upon the $O(r^2)$ projector
$(\partial x^i/\partial x^b) (\partial x^j/\partial x^c)$ also yields
Eq.~(\ref{projectedh}).

Let us obtain the asymptotic expansion for the action of $\Bderiv$ on an 
$s$ spin--weighted scalar. Using Eqs.~(\ref{MandM}) in the appendix 
Eq.~(\ref{Sigmaframe}b) for the dyad leg $e_{\hat{3}}$ or via simple
inference, one finds the not unexpected result,
\begin{equation}
m^{b} = \mu^b -
\half\left(
a_{\mu\mu} \bar{\mu}{}^{b} + a_{\mu\bar{\mu}} \mu^b
\right)r^{-1}
+ O(r^{-2-\varepsilon})\, ,
\label{mexpansion}
\end{equation}
where, for example, $\mu^a = f^{ab} \mu_b = O(1/r)$. 
Now define the operator $\delta = m^a \partial/\partial x^a$, and let 
$\delta_0 = r \mu^a \partial/\partial x^a$ be the corresponding operator
on the unit sphere. On scalars, these operators agree with the
ones discussed in {\em Appendix B.1}. Contraction of 
Eq.~(\ref{mexpansion}) on $\partial/\partial x^b$ then gives
\begin{equation}
\delta \sim \delta_0 r^{-1} -  
\half\left(
a_{\mu\mu} \bar{\delta}_0 + a_{\mu\bar{\mu}} \delta_0
\right)r^{-2}
\, .
\label{deltaexpansion}
\end{equation}
Next, we insert the co--frame expansions (\ref{MandM}) into the appendix
result (\ref{Sigmaconnection}b) for the $B$ connection form $\omega
= \omega_{\hat{4}\hat{3}\hat{3}}$, thereby reaching
\begin{equation}
\omega \sim \omega_0r^{-1} +
\half\left(\delta_0[a_{\mu\bar{\mu}}] 
- \bar{\delta}_0[a_{\mu\mu}] 
- \omega_0 a_{\mu\bar{\mu}}
- \bar{\omega}_0 a_{\mu\mu}\right)r^{-2}
\, ,
\label{omegaexpansion}
\end{equation}
where $\omega_{0} = - \bar{\zeta}/\sqrt{2}$ is the connection form on the 
unit sphere.
Now, by definition $\Bderiv = \delta - s \omega$, whence we find
\begin{equation}
\Bderiv \sim \Bderiv_{0} r^{-1}
- \half\left[
a_{\mu\bar{\mu}} \Bderiv_0
+ a_{\mu\mu} \bar{\Bderiv}_0
+ s(\Bderiv_0 a_{\mu\bar{\mu}})
- s(\bar{\Bderiv}_0 a_{\mu\mu})\right] r^{-2}
\label{ethexpansion}
\end{equation}
as the sought--for $\Bderiv$ expansion.
To get the corresponding expansion for $\bar{\Bderiv}$, simply complex
conjugate (\ref{ethexpansion}) and then send $s \rightarrow - s$.

Let us obtain expansions for the dyad components of the $B$ extrinsic
curvature tensor. As a connection form, the trace is given by the formula
$k = 2k_{\hat{3}\hat{4}} = - 2\omega_{\hat{3}\vdash\hat{4}}$. So we insert 
the expansions (\ref{MandW}), (\ref{MandM}), and (\ref{ethexpansion}) into 
appendix formula (\ref{Sigmaconnection}c) for 
$\omega_{\hat{3}\vdash\hat{4}}$, obtaining
\begin{equation}
k \sim -2r^{-1}
+ (
a_{\nu\nu} +  a_{\mu\bar{\mu}}
+ \Bderiv_0 a_{\nu\bar{\mu}}
+ \bar{\Bderiv}_0 a_{\nu\mu})r^{-2}
\label{expansionfork}
\end{equation}  
as the desired expansion.
Similarly, starting with $k_{\hat{3}\hat{3}} 
= -\omega_{\hat{3}\vdash\hat{3}}$ and Eq.~(\ref{Sigmaconnection}d),
we get 
\begin{equation}
k_{mm} \sim (
\half a_{\mu\mu}
+ \Bderiv_0 a_{\nu\mu})r^{-2}
\label{expansionforkmm}
\end{equation}
as the expansion for the trace--free part of $k_{ab}$. 

Finally, let us obtain the asymptotic expansion for the scalar curvature
${\cal R}$ of $B$. Into the appendix formula (\ref{Bcurvature}) for 
${\cal R}$, we insert the expansions (\ref{deltaexpansion}) and 
(\ref{omegaexpansion}), with result
\begin{equation}
{\cal R} \sim 2 r^{-2}
+ (\Bderiv^2_0 a_{\bar{\mu}\bar{\mu}}
+ \bar{\Bderiv}{}^2_0 a_{\mu\mu}
- 2\bar{\Bderiv}_0\Bderiv_0 a_{\mu\bar{\mu}}
- 2 a_{\mu\bar{\mu}})r^{-3}\, .
\label{Rexpansion}
\end{equation}

Our expansions for $k = 2k_{m\bar{m}}$, 
$k_{mm}$, $\Bderiv$, and ${\cal R}$ determine
the leading order behavior in radial expansions for some
components of the $\Sigma$ Riemann tensor. Indeed from the 
Gau{\ss}-Codazzi-Mainardi embedding equations, 
\cite{Yorkaction,BLYss}
\begin{subequations}\label{newembed12}
\begin{eqnarray}
      (k_{m\bar{m}}){}^2
     - k_{mm} k_{\bar{m}\bar{m}}
     - {\textstyle \frac{1}{2}}{\cal R}  & = &
       R_{m\bar{m}m\bar{m}}
\\
       \bar{\Bderiv} k_{mm}
     - \Bderiv k_{m\bar{m}} & = &
       R_{\vdash m\bar{m}m}\, ,
\end{eqnarray}
\end{subequations}
we may infer the explicit expressions for the coefficients
${}^{3} R_{m\bar{m}m\bar{m}}$ and ${}^{3} R_{\vdash m\bar{m}m}$
in the expansions
\begin{subequations}\label{sigRieexps}
\begin{eqnarray}
       R_{m\bar{m}m\bar{m}} & \sim &
       {}^{3} R_{m\bar{m}m\bar{m}} r^{-3}
\\
       R_{\vdash m\bar{m}m} & \sim &
       {}^{3} R_{\vdash m\bar{m}m} r^{-3}\, .
\end{eqnarray}
\end{subequations}
Eq.~(\ref{newembed12}a) determines that
\begin{equation} 
{}^3 R_{m\bar{m}m\bar{m}} = - \half {}^3{\cal R} - {}^2 k\, ,
\label{RRandk}
\end{equation}
whence we obtain
\begin{equation}
{}^3 R_{m\bar{m}m\bar{m}} = - a_{\nu\nu}   
-\Bderiv_{0} a_{\nu\bar{\mu}}
-\bar{\Bderiv}_{0} a_{\nu\mu}
-\half\Bderiv^2_0 a_{\bar{\mu}\bar{\mu}}
-\half\bar{\Bderiv}{}^2_0 a_{\mu\mu}
+\bar{\Bderiv}_0\Bderiv_0 a_{\mu\bar{\mu}}\, .
\label{secfour3curvature}
\end{equation}
Likewise, Eq.~(\ref{newembed12}b) determines that
\begin{equation}
{}^{3} R_{\vdash m\bar{m}m} 
= \bar{\Bderiv}_0 {}^2 k_{mm} - \Bderiv_0 {}^2 k_{m\bar{m}}\, ,
\end{equation}
whence 
\begin{equation}
{}^{3} R_{\vdash m\bar{m}m} = 
\half a_{\nu\mu}
-\half\Bderiv_0                       
a_{\nu\nu}
+ \half \bar{\Bderiv}_0 a_{\mu\mu}
-\half\Bderiv_0
 a_{\mu\bar{\mu}}
-\half\Bderiv^2_0 a_{\nu\bar{\mu}}
+\half\bar{\Bderiv}_0
\Bderiv_0 a_{\nu\mu}\, .
\end{equation}
To reach the last equation, we have appealed to the commutator equation
$2[\bar{\Bderiv}_0,\Bderiv_0] a_{\nu\mu} = a_{\nu\mu}$ discussed in
{\em Appendix B.1}.

The following consistency check affords some confidence in our results. 
First, show that
\begin{equation}
\sqrt{\sigma} = {\rm i}(|m_{\bar{\zeta}}|^2 - |m_{\zeta}|^2)
= {\rm i}2r^2 P^{-2}
\left[1 + a_{\mu\bar{\mu}}r^{-1}
+ O(r^{-1-\varepsilon})
\right]\, ,
\label{detsigma}
\end{equation}
where the ${\rm i}$ is necessary for $d^2 x \sqrt{\sigma}$ to be 
real as $d\zeta\wedge d\bar{\zeta}$ is pure imaginary.  Now the area 
form on a radius--$r$ round sphere is
\begin{equation}
{\rm i} \bar{\mu}_a \mu_b\, dx^a \wedge dx^b 
= {\rm i}2 r^2 P^{-2} d\zeta\wedge d\bar{\zeta}\, ;
\end{equation}
and with this fact and Eq.~(\ref{Rexpansion}), we perform a direct 
calculation showing that
\begin{equation}
\int_{B}d^2 x \sqrt{\sigma} {\cal R} = 8\pi +
O(r^{-1-\varepsilon})\, .
\end{equation}
That is to say, the Gau{\ss}--Bonnet Theorem holds to the same level
of accuracy as our approximations thus far. We point out that the
presence of the sole term in ${}^3{\cal R}$ which is not a 
unit--sphere divergence, namely $-2 a_{\mu\bar{\mu}}$, plays a crucial 
role in this agreement, as it cancels a similar such term in
$\sqrt{\sigma}$. Note that our discussion here has also 
established the last section's Eq.~(\ref{elementofB}) as well as
Eq.~(\ref{secondresult}) at the beginning of this section. Do note,
however, that the coordinates used here are $(\zeta,\bar{\zeta})$,
whereas in {\em Section III} the $B$ coordinates used were
$(\theta,\phi)$. These two systems are related by a non--trivial (in 
fact imaginary) Jacobian. Therefore, $d^2x$ here is not the $d^2x$ 
from {\em Section III}, although of course $d^2x\sqrt{\sigma}$ is the
same here as there.

\subsection{Parity}

Define the action of the parity operator ${\cal P}$ via ${\cal P}(x^k)
= - x^k$, whence it follows that ${\cal P}(\nu^k) = -\nu^k$. Moreover,
one can show that ${\cal P}(\zeta) = -1/\bar{\zeta}$, and from this
result and (\ref{muvector}) that 
\begin{equation}
{\cal P}(\mu^i) = -(\bar{\zeta}/\zeta) \bar{\mu}{}^i\, .
\end{equation}
We can then immediately write
\begin{subequations} \label{aparity}
\begin{eqnarray}
{\cal P}(a_{\nu\nu}) = a_{\nu\nu}\, , & \qquad &
{\cal P}(a_{\nu\mu}) = (\bar{\zeta}/\zeta) a_{\nu\bar{\mu}}\, ,
\\
{\cal P}(a_{\mu\bar{\mu}}) = a_{\mu\bar{\mu}}\, ,
& & 
{\cal P}(a_{\mu\mu}) = 
(\bar{\zeta}/\zeta)^2 a_{\bar{\mu}\bar{\mu}}\, .
\end{eqnarray}
\end{subequations}
Moreover,
\begin{equation}
{\cal P}(\delta_{0}) = (\bar{\zeta}/\zeta)\bar{\delta}_0
\, ,
\qquad
{\cal P}(\omega_{0}) = -(\zeta)^{-2}\bar{\omega}_0\, ,
\label{deltaparity}
\end{equation}
where the first of these formulae follows since
${\cal P}(\delta_0) = {\cal P} (r\mu^i\partial/\partial x^i)$. Note
that the minus sign difference between the formulae for
${\cal P}(\delta_0)$ and ${\cal P}(\mu^i)$ stems from
${\cal P}(\partial/\partial x^i) = -\partial/\partial x^i$.
With the formulae amassed so far, it is fairly easy to establish the
following:
\begin{subequations}\label{ethparity}
\begin{eqnarray}
{\cal P}(\Bderiv^2_0 a_{\bar{\mu}\bar{\mu}}) & = & 
\bar{\Bderiv}{}^2_0 a_{\mu\mu} 
\\
{\cal P}(\Bderiv_{0}a_{\nu\bar{\mu}}) & = &
\bar{\Bderiv}_{0}a_{\nu\mu}
\\
{\cal P}(\Bderiv_0 a_{\nu\mu}) & = & (\bar{\zeta}/\zeta)^2
\bar{\Bderiv}_0 a_{\nu\bar{\mu}}
\\
{\cal P}(\bar{\Bderiv}_0\Bderiv_0 a_{\mu\bar{\mu}}) & = &
\bar{\Bderiv}_0\Bderiv_0 a_{\mu\bar{\mu}}\, .
\end{eqnarray}
\end{subequations}
These results and their complex conjugates in turn show that 
${}^3{\cal R}$, ${}^2 k$, ${}^3 R_{m\bar{m}m\bar{m}}$,
and ${}^2 k_{mm} {}^2 k_{\bar{m}\bar{m}}$ are all of even parity.

\subsection{Energy integral}
Let us return to the surface integral (\ref{BYintegral}) and 
total energy scenario. Consider 
the expression (\ref{RRandk}) derived from the 
Gau\ss--Codazzi--Mainardi equations. A similar
formula is associated with the isometric embedding of $B$ in 
$\barSigma$ and the {\em Ansatz}
\begin{equation}
k|^{\rm ref} \sim -2r^{-1} + {}^2k|^{\rm ref} r^{-2}\, .
\label{secfouransatz}
\end{equation}
Namely,
\begin{equation}
         {}^2k|^{\rm ref} =
         - {\textstyle \frac{1}{2}}\,{}^{3}{\cal R}\, .
\label{twokref}
\end{equation}
We give a more careful derivation of this result in the next 
subsection. Therefore, appealing to Eq.~(\ref{RRandk}) we get 
\begin{equation}
k - k|^{\rm ref} \sim - {}^{3}\! R_{m\bar{m}m\bar{m}}r^{-2}\, ,
\end{equation}
and, noting that 
\begin{equation}
2R_{m\bar{m}m\bar{m}} = 
-\sigma^{ij}\sigma^{kl} R_{ikjl}\, ,
\label{RiemanntoRiemann}
\end{equation}
we may also write this as
\begin{equation}
k - k|^{\rm ref} \sim\,  {}^{3}\! 
\left[\half\sigma^{ij}\sigma^{kl} R_{ikjl}
\right]r^{-2}\, .
\end{equation}
Moreover, for the $\Sigma$ components of the spacetime Riemann 
tensor, one has \cite{Yorkaction}
\begin{equation}
\Re_{ijkl} = R_{ijkl} + K_{ik} K_{jl} - K_{il} K_{jk}\, .
\label{ReRKK}
\end{equation}
Since $K_{ij} = O(r^{-2})$, the terms quadratic in $K_{ij}$ are 
in fact $O(r^{-4})$ terms, whence we have established 
Eq.~(\ref{firstresult}).

\subsection{Center--of--mass integral}

When considering the surface integral (\ref{BYintegral}) in the 
center--of--mass scenario, we keep more powers of inverse radius
in the $k$ expansion than we kept in Eq.~(\ref{expansionfork}). 
We now write 
\begin{equation}
       k \sim - 2 r^{-1} 
              + {}^2 k r^{-2} 
              + {}^{(2+\varepsilon)} k r^{-2-\varepsilon} 
              + {}^3 k r^{-3}\, ,
\end{equation}
with the previously given expression for ${}^2 k$ still valid. From the 
embedding equation (\ref{newembed12}a) we infer that
\begin{subequations}\label{kcalRRiemann}
\begin{eqnarray}
       2\, {}^{(2+\varepsilon)}k_{m\bar{m}} & = &
     - {\textstyle \frac{1}{2}} {}^{(3+\varepsilon)}{\cal R}
     - {}^{(3+\varepsilon)} R_{m\bar{m}m\bar{m}}
       \label{kcalRRiemanna}
       \\
       2\, {}^{3}k_{m\bar{m}}               & = &
       \big({}^{2}k_{m\bar{m}}\big){}^{2}   
     - {}^{2}k_{mm}{}^{2}k_{\bar{m}\bar{m}}
     - {\textstyle \frac{1}{2}} {}^{4}{\cal R}
     - {}^{4} R_{m\bar{m}m\bar{m}}\, ,
       \label{kcalRRiemannb}
\end{eqnarray}
\end{subequations}
and so ---with the results of the parity subsection--- we have  
\begin{eqnarray}
k & \sim & -2r^{-1} + \big({\rm E.P.T.}\big)r^{-2}
     \nonumber \\
  & & - \big({\textstyle \frac{1}{2}}
        {}^{(3+\varepsilon)}{\cal R}
      + {}^{(3+\varepsilon)} 
        R_{m\bar{m}m\bar{m}}\big)r^{-2-\varepsilon}
      - \big({\rm E.P.T.} 
      + {\textstyle \frac{1}{2}}
        {}^{4}{\cal R}
      + {}^{4} R_{m\bar{m}m\bar{m}}\big)r^{-3}\, .
\end{eqnarray}
Again, E.P.T. stands for generic even parity terms. We claim that
\begin{equation}
k|^{\rm ref} \sim -2r^{-1} + \big({\rm E.P.T.}\big)r^{-2}
       - \big({\textstyle \frac{1}{2}}
        {}^{(3+\varepsilon)}{\cal R}\big)r^{-2-\varepsilon}         
       - \big({\rm E.P.T.} + {\textstyle \frac{1}{2}}
       {}^{4}{\cal R}\big)r^{-3}\, ,
\label{showthis}
\end{equation}
and will establish this result below. We note in
passing that the expansion (\ref{showthis}) agrees with the ``lightcone
reference'' $-\sqrt{2\cal R}$ derived in Ref.~\cite{Lau2}. Combination
of these equations with Eq.~(\ref{RiemanntoRiemann}) gives
\begin{equation}
       k - k|^{\rm ref} \sim   \big({\rm E.P.T.}\big)r^{-2}
                             + {}^{(3+\varepsilon)}\!
                               \left[\half
                               \sigma^{ij}\sigma^{kl} 
                               R_{ikjl}
                               \right]r^{-2-\varepsilon}
                             + \big({\rm E.P.T.} 
                             + {}^{4}\!
                               \left[\half
                               \sigma^{ij}\sigma^{kl} 
                               R_{ikjl}
                               \right]\big)r^{-3}\, .
\label{kminuskrefexp}
\end{equation}
Now, since $K_{ij} \sim d_{ij}(\nu^k)r^{-2}$ with $d_{ij}$ of odd parity,
the quadratic extrinsic curvature terms in (\ref{ReRKK}) are of 
leading $r^{-4}$ order and even parity. Hence, we have
\begin{subequations}\label{someRRresults}
\begin{eqnarray}
{}^{(3+\varepsilon)}\!
\left[\half\sigma^{\mu\nu}\sigma^{\lambda\kappa} 
\Re_{\mu\lambda\nu\kappa}
\right] & = & {}^{(3+\varepsilon)}\! \left[\half\sigma^{ij}\sigma^{kl} 
R_{ikjl}
\right]
\label{someRRresultsa} \\
{}^{4}\!
\left[\half\sigma^{\mu\nu}\sigma^{\lambda\kappa} \Re_{\mu\lambda\nu\kappa}
\right] & = & {}^{4}\! \left[\half\sigma^{ij}\sigma^{kl} R_{ikjl}
\right] + {\rm E.P.T.}\, ,
\label{someRRresultsb}
\end{eqnarray}
\end{subequations}
and this result along with (\ref{kminuskrefexp}) establishes 
Eq.~(\ref{thirdresult}).

Let us now verify (\ref{showthis}). We must consider the
Gau{\ss}--Codazzi--Mainardi equations associated with the isometric 
embedding of $B$ in $\barSigma \simeq E^{3}$. Using 
$\bark_{ab} = (k|^{\rm ref})_{ab}$ as a shorthand, 
these are
\begin{subequations}\label{newembed34}
\begin{eqnarray}
      (\bark_{m\bar{m}}){}^2
     - \bark_{mm}
       \bark_{\bar{m}\bar{m}}
     - {\textstyle \frac{1}{2}}{\cal R}  & = & 0
\\
       \bar{\Bderiv} \bark_{mm}
     - \Bderiv \bark_{m\bar{m}}
                            & = & 0\, .
\end{eqnarray}
\end{subequations}
We solve these equations ---in the sense of asymptotic expansions--- for 
$\bark_{m\bar{m}}$ and $\bark_{mm}$, viewing the intrinsic $B$
geometry, both $\Bderiv$ and ${\cal R}$, as fixed. We make the following 
{\em Ans\"{a}tze}:
\begin{subequations}\label{ansatz2}
\begin{eqnarray}
       \bark_{m\bar{m}} & \sim &
       - r^{-1}         
       + {}^{2}\bark_{m\bar{m}} r^{-2}
       + {}^{(2+\varepsilon)}\bark_{m\bar{m}} r^{-2-\varepsilon}
       + {}^{3}\bark_{m\bar{m}} r^{-3} 
\\
       \bark_{mm}       & \sim & 
       {}^2 \bark_{mm} r^{-2}
\, . 
\end{eqnarray}
\end{subequations}
We remark that consistency of the {\em Ans\"{a}tze} adopted here
follows from the fact that ---as shown in {\em Section III}--- the 
non--trivial flat background metric $\barh_{ij}$ admits an expansion 
analogous to $h_{ij}$. Therefore, one expects all of the asymptotic 
expansion calculated in {\em Section IV.A} to carry over for 
$\barh_{ij}$. However, here we endeavor to follow an independent 
approach. From Eqs.~(\ref{ansatz2}) we first algebraically find 
that
\begin{subequations}\label{eqagain}  
\begin{eqnarray}
{}^2 \bark_{m\bar{m}} & = & 
-{\textstyle\frac{1}{4}} {}^3 {\cal R} 
\label{eqagaina}\\
{}^{(2+\varepsilon)}\bark_{m\bar{m}} & = & 
-{\textstyle\frac{1}{4}} {}^{(3+\varepsilon)}{\cal R}\, ,
\end{eqnarray}
\end{subequations}
establishing in particular the claim made earlier in 
Eq.~(\ref{twokref}). Clearly then ---as the previous 
parity analysis shows--- ${\cal P}({}^2 \bark_{m\bar{m}}) = 
{}^2 \bark_{m\bar{m}}$, which justifies the E.P.T. term next
to the $r^{-2}$ on the {\sc rhs} of Eq.~(\ref{showthis}). Also 
algebraically, we obtain
\begin{eqnarray}
       2\,{}^{3}\bark_{m\bar{m}} & = &   
       \big({}^{2}\bark_{m\bar{m}}\big){}^{2}
     - {}^{2}\bark_{mm}{}^{2}\bark_{\bar{m}\bar{m}}
     - {\textstyle \frac{1}{2}}{}^{4}{\cal R}\, ,
\end{eqnarray}       
where ---as seen from Eqs.~(\ref{newembed34}b) and (\ref{eqagaina})--- 
the weight--two scalar ${}^2\bark_{mm}$ solves the {\sc pde}
\begin{equation}
       \bar{\Bderiv}_{0}\, {}^{2}\bark_{mm} =
         - {\textstyle \frac{1}{4}} \Bderiv_{0}
           {}^{3}{\cal R}\, .
\label{pde}
\end{equation}
{\sc Lemma.} ${\cal P}({}^2\bark_{mm}) = 
(\bar{\zeta}/\zeta)^2 ({}^2\bark_{\bar{m}\bar{m}}$), whence
${}^{2}\bark_{mm}{}^{2}\bark_{\bar{m}\bar{m}}$ is of even parity
[which is the remaining needed piece to verify Eq.~(\ref{showthis})].
\\[2mm]
To start the proof, let us show that there exists a unique solution 
${}^2\bark_{mm}$ to the {\sc pde} given in Eq.~(\ref{pde}), where we
view the {\sc rhs} as a prescribed source determined by 
Eq.~(\ref{Rexpansion}). As is well--known, for an 
equation of the form $\bar{\Bderiv}_0 f = g$ with $g$ a prescribed 
weight--one source, a unique inverse $\bar{\Bderiv}{}^{-1}_0$ to the 
operation $\bar{\Bderiv}_0$ exists if 
and only if \cite{ERP}
\begin{equation}
\oint d\Omega\,
{}_1\overline{Y}_{1m}\, g = 0\, ,
\end{equation}
where again $\oint d\Omega$ denotes average
over the unit sphere $S^2$ and the ${}_1 Y_{lm}$ are spin--$1$ spherical
harmonics. Using Eq.~(\ref{conjugate}) from {\em Appendix B.1}, we see 
that for our equation the issue at hand is whether or not
\begin{equation}
\oint d\Omega\,  {}_{-1}Y_{1m}\,
\Bderiv_{0}{}^{3}{\cal R}
\label{integralYdR}
\end{equation}
vanishes for $m = -1,0,1$. Simple integration by parts will not show that
the integral vanishes, since $\Bderiv_{0} ({}_{-1}Y_{1m}) \neq 0$. 
However, for any scalar curvature ${\cal R}$, there exists a weight--two
scalar $Q$ such that $\Bderiv {\cal R} = \bar{\Bderiv} Q$. \cite{Tod}
More precisely, letting $\psi$ denote the conformal factor relating 
$\sigma_{ab}$ to the line--element $ds^2_0$ of the unit sphere, as in
\begin{equation}
ds^2_0 = \psi^2 \sigma_{ab}dx^a dx^b\, ,
\end{equation}
Tod shows that \cite{Tod}
\begin{equation}
Q = 4 \Bderiv^2 \log\psi - 4(\Bderiv \log \psi)^2\, .
\label{Qequation}
\end{equation}  
Assume $\psi \sim r^{-1} + {}^2\psi r^{-2}$ and that ${}^2\psi$ is of
even parity. Since the $\log$
terms are all differentiated we may write
\begin{eqnarray}
\log\psi & = & -\log r + \log\left[1 + {}^2\psi r^{-1}
+ O(r^{-1-\varepsilon})\right]
\nonumber \\ 
& = &  -\log r + {}^2\psi r^{-1} +  
O(r^{-1-\varepsilon})\, ,
\end{eqnarray}
whence (\ref{Qequation}) in tandem with the expansion 
(\ref{ethexpansion}) for $\Bderiv$ leads to
\begin{equation}
Q \sim 4 \Bderiv^2_0 ({}^2\psi)r^{-3}\, .
\end{equation}
This is a consistent result since 
$\bar{\Bderiv}Q = \Bderiv R = O(r^{-4})$, the leading
$2r^{-2}$ term in ${\cal R}$ being annihilated by the angular
derivatives in $\Bderiv$. We have then the leading-order identity 
$\Bderiv_0 {}^3 {\cal R} = 4\bar{\Bderiv}_0 \Bderiv^2_0 ({}^2\psi)$,
and so ---dropping a factor of 4--- the integral (\ref{integralYdR}) 
just above becomes
\begin{equation}
\oint d\Omega\,  {}_{-1}Y_{1m}\,
\bar{\Bderiv}_0
\Bderiv^2_0 ({}^2\psi) = 0\, .
\end{equation}
The last equality follows by integration by parts and 
$\bar{\Bderiv}_{0} ({}_{-1} Y_{1,m}) = 0$. Thus, the uniqueness of
$\bar{\Bderiv}^{-1}$ is established. Whence we have
\begin{eqnarray}
{}^2\bark_{mm} & = & 
-{\textstyle \frac{1}{4}}
\bar{\Bderiv}^{-1}_0 \Bderiv_0 {}^3{\cal R}
\nonumber \\
& = & -{\textstyle \frac{1}{4}}\bar{\Bderiv}^{-1}_0 
\big[4\bar{\Bderiv}_0 \Bderiv^2_0 ({}^2\psi)\big]
\nonumber \\
& = & -\Bderiv^2_0 ({}^2\psi)\, .
\end{eqnarray}
Since ${}^{2}\psi$ is an even parity scalar function, the lemma 
then follows by simple calculations as outlined in the parity
subsection.

\section{Concluding remarks}
Our investigation has shown the mass--aspect to be the following: 
[cf.~Eqs.~(\ref{averageR3},\ref{secfour3curvature},
\ref{RiemanntoRiemann},\ref{ReRKK})]
\begin{equation}
{\textstyle \frac{1}{4}}\,
{}^3\!\big[\sigma^{\mu\nu}\sigma^{\lambda\kappa}
\Re_{\mu\lambda\nu\kappa}\big] 
= {\textstyle \frac{1}{4}}\big(2 a_{\nu\nu}
+2\Bderiv_{0} a_{\nu\bar{\mu}}
+2\bar{\Bderiv}_{0} a_{\nu\mu}
+\Bderiv^2_0 a_{\bar{\mu}\bar{\mu}}
+\bar{\Bderiv}{}^2_0 a_{\mu\mu}
-2\bar{\Bderiv}_0\Bderiv_0 a_{\mu\bar{\mu}}\big)\, .
\label{CRmassaspect}
\end{equation}
[Upon {\em proper} averaging over the unit sphere, 
that is to say $(4\pi)^{-1}\oint d\Omega$ integration, 
the mass--aspect yields the total energy.] This mass--aspect may be 
compared with the one stemming from the {\sc adm} integral 
(\ref{ADMenergy}). Namely,
\begin{equation}
{\textstyle \frac{1}{4}}\,
{}^2\!\big[n^kh^{ij}(\partial_j h_{ik}
-\partial_k h_{ij})\big]
= {\textstyle \frac{1}{4}} 
\big(2 a_{\nu\nu} +\Bderiv_0 a_{\nu\bar{\mu}}
+\bar{\Bderiv}_0 a_{\nu\mu}\big)\, .
\label{CADMmassaspect}
\end{equation}
(\ref{CRmassaspect}) and (\ref{CADMmassaspect}) differ by a pure 
divergence on the unit sphere. We note that the reference term 
$k|^{\rm ref}$ in Eqs.~(\ref{BYintegral},\ref{secfouransatz})
plays a crucial role in yielding a 
mass--aspect (\ref{CRmassaspect}) whose proper unit--sphere average 
agrees with the {\sc adm} energy. Indeed, not only does $k|^{\rm ref}$ 
remove the leading order divergent contribution to the ``unreferenced
energy'' [the proper $B$ integral of $(8\pi)^{-1}k$ alone
which blows up like $r$], at the next 
order it removes a dangerous factor of $a_{\mu\bar{\mu}}$ from ${}^2k$ 
in Eq.~(\ref{expansionfork}). Indeed, the total energy 
$(8\pi)^{-1}\oint d\Omega\, a_{\nu\nu}$ is {\em not} 
equal to $(8\pi)^{-1}\oint d\Omega\, {}^2 k$ in general.

In stark contrast with the energy scenario, we point out that the 
results of {\em Appendix C.1} (in particular the lemmas as they pertain 
to coefficients of the ${\cal R}$ expansion) show that the $k|^{\rm ref}$ 
term in Eqs.~(\ref{BYintegral},\ref{showthis}) in fact makes no 
contribution at all to the $r\rightarrow\infty$ limit 
in the center--of--mass scenario! This suggests that the \bom integral 
might be compared directly to the proper $B$ integral of $(8\pi)^{-1}N k$ 
as an alternate way of checking the correspondence between $H_B$ and 
$M^\perp_B$ in the $r\rightarrow\infty$ limit, and one which bypasses 
the issue of the reference term and the solution to the embedding 
equations altogether. (Using {\em Section IV} techniques, we have 
performed such a check. 
The calculation amounts to a tedious exercise in perturbation theory.) 
However, this alternate way requires that one somehow know in advance 
that the reference term makes no contribution to the center--of--mass 
limit, so the reasoning would seem circular. Furthermore, we note that 
our asymptotic solution to the embedding equations presented in 
{\em Appendix A.2} and the analysis of {\em Section III} justify the 
{\em Ans\"{a}tze} (\ref{ansatz2}) of {\em Section IV}, ultimately 
showing that we have needed to use this solution and that analysis in 
verifying that the reference term does not contribute to the center of 
mass after all. By way of comparison with the last sentence of the
preceding paragraph, we mention that 
---knowing the reference term makes no
contribution--- the $k$th center--of--mass Cartesian component is 
indeed $(8\pi)^{-1}\oint d\Omega\,{}^3 k\nu^k$, and we find
$(8\pi)^{-1}\oint d\Omega\, 
(c_{\nu\nu} 
+ 2 c_{\mu\bar{\mu}} 
+ \Bderiv_0 c_{\nu\bar{\mu}}
+ \bar{\Bderiv}_0 c_{\nu\mu}) 
\nu^k$
as the explicit expression for this component. Moreover,
under changes in $c_{ij}$ induced by coordinate transformations
on $h_{ij}$ of the form (\ref{expansion-X},\ref{expansion-X/x}), 
we find that this unit--sphere integral is an invariant.

These considerations also highlight the {\em difference} between the 
Brown--York and \bom integrals. Indeed, 
Eqs.~(\ref{firstresult},\ref{kminuskrefexp},\ref{someRRresults})
show {\em explicitly} that even the {\em integrand} of $H_B$ in 
(\ref{BYintegral}) vanishes to a high order in $1/r$ for trivial 
initial data. Actually, of course, by definition $k-k|^{\rm ref}$ 
is identically zero to all orders if $h_{ij}$ is a Euclidean metric
and $\Sigma$ is $E^3$. However, as we have seen in 
{\em Section III.F}, the \bom {\em integrand} can be 
non--zero even for trivial data, in which case the vanishing of 
the full integral relies on the integration itself. This difference may 
well be relevant when
supertranslations are brought into play, and we hope that our
extremely careful treatment of the reference term will prove useful in
future investigations.

\section{acknowledgments}
Both SRL and ANP wish to extend special thanks to Professor
J.~M.~Nester for encouragement and for organizing the Workshop on 
Geometric Physics held 24--26 July 2000 at the Center for 
Theoretical Studies, National Tsing Hua University, Hsinchu, 
TAIWAN. Work on this paper began shortly after the Workshop.
DB and ANP also thank Professor J.~M.~Nester for useful discussions
about this work during the Conference ICGA--5 in Moscow in 
October 2001. SRL is grateful for discussions with Professor 
Y.~J.~Ng about canonical boost generators in special 
relativistic field theory, and he thanks Moscow State 
University for hospitality during a visit in July 2002.
Finally, all authors wish to thank both 
Prof.~L.~B.~Szabados and Prof.~J.~W.~York for stimulating 
correspondence and remarks on the manuscript.

\noindent
{\em Note added in proof:} As we have often noted, our
calculations have been carried out with our more detailed
version (\ref{BOM}) of 
Beig--\'{o} Murchadha fall--off \cite{BeigOMurchadha}. 
Recently, Szabados has carefully analyzed the Poincare structure of
asymptotically flat spacetimes and found Beig--\'{o} Murchadha
fall--off to be the weakest possible fall--off which ensures
finiteness of the angular momentum and center of 
mass \cite{SzabadosPoinc}.

\appendix

\section{Key identities and embedding equations}
\subsection{Derivation of key identities}
Let us now establish the main identities given in
Eqs.~(\ref{identity1},\ref{identity2}).
We begin by collecting some preliminary identities needed to get the
main ones. Subtracting the two equations in (\ref{projections}), we find
\begin{equation}
\Delta \sigma^{ik} = \Delta q^{ik} - \barn^{i} \Delta n^k
                  - \Delta n^i \barn^k
                  - \Delta n^i \Delta n^k\, ,
\label{idone}
\end{equation}
which upon contraction with $\barn_i \barn_k$ yields
\begin{equation}
\barn_{i} \Delta n^i =     \half \barn_i \barn_k\left(
                            \Delta q^{ik}
                          - \Delta\sigma^{ik}\right)
                          - \half\left(\barn_i \Delta n^i\right){}^2\, .
\label{idtwo}
\end{equation}
Next, contracting Eq.~(\ref{idone}) on $\half\barsigma_{ik}$ we get
\begin{equation}
\half\barsigma_{ik}\left(\Delta q^{ik} - \Delta\sigma^{ik} - \Delta n^i
\Delta n^k\right) = 0\, ,
\label{idthree}
\end{equation}
and so addition of (\ref{idthree}) to the {\sc rhs} of (\ref{idtwo}) gives
\begin{equation}
\barn_{i} \Delta n^i =     \half \barh_{ik}\left(
                            \Delta q^{ik}
                          - \Delta\sigma^{ik}\right)
                          - \half\barh_{ik}\Delta n^i\Delta n^k\, .
\label{idfour}
\end{equation}
Equating the righthand sides of Eqs.~(\ref{idtwo},\ref{idfour}), we then
have
\begin{equation}
\barn_i\barn_k\left(\Delta q^{ik}-\Delta\sigma^{ik}\right)
= \barh_{ik}\left(\Delta q^{ik}-\Delta\sigma^{ik}\right) +
\left(\barn_i \Delta n^i\right){}^2 - \barh_{ik}\Delta n^i\Delta n^k\, .
\label{idfive}
\end{equation}
Finally, we contract (\ref{idone}) on $\barn_i$ and insert Eq.~(\ref{idtwo})
into the result, thereby obtaining
\begin{eqnarray}
\Delta n^k    & =  &
              \barn_i \left(\Delta q^{ik}-\Delta\sigma^{ik}\right)
            - \half \barn^k \barn_i\barn_j
              \left(\Delta q^{ij} - \Delta\sigma^{ij}\right)
             \nonumber \\
 & &         + \half \barn^k \left(\barn_i \Delta n^i\right){}^2
              -\barn_i \Delta n^i\Delta n^k\, .
\label{idsix}
\end{eqnarray}

Let us now turn to the expression $(k-\bark)$. Straightaway, we have
\begin{equation}
k - \bark = -\barD_i \Delta n^i - \barn^i\Delta\Gamma^l{}_{li}
                                -  \Delta n^i\Delta\Gamma^l{}_{li}\, ,
\end{equation}
where the difference in Christoffel symbols is $\Delta \Gamma^l{}_{li}
= \Gamma^l{}_{li} - \barGamma^l{}_{li}$. Now, standard formulae show
both that $\Delta \Gamma^l{}_{li} = \half h^{kl}\barD_i\Delta h_{kl}$
and $h^{kl}\barD_k \Delta h_{il} = - h_{il}\barD_k \Delta q^{kl} $, and
with these we rewrite the last equation as
\begin{eqnarray}
   k - \bark & = & \half\barn^i h^{kl}
               \left(\barD_k \Delta h_{il} -
               \barD_i\Delta h_{kl}\right)
             - \barD_i \Delta n^i \nonumber \\
             &   &
             + \half \barn^i h_{il}\barD_k\Delta q^{kl}
             - \half h^{kl}\Delta n^i \barD_i \Delta h_{kl}\, .
\label{step}
\end{eqnarray}
Focus attention on the next to last term in this equation, which we
rewrite as
\begin{equation}
\half \barn^i h_{il}\barD_k\Delta q^{kl}
= \half \barn^i \Delta h_{il}\barD_k\Delta q^{kl}
+ \half \barn_{l}\barD_k\Delta q^{kl}   \, .
\end{equation}
An appeal to Eq.~(\ref{idone}) then shows that
\begin{eqnarray}
\half \barn^i h_{il}\barD_k\Delta q^{kl}
& = & \half \barn^i \Delta h_{il}\barD_k\Delta q^{kl}
+ \half \barn_{i}\barD_k\Delta\sigma^{ik}
\nonumber \\
& & + \half \barn_{i}\barD_k \left( \Delta n^i \Delta n^k\right)
+ \barn_{i}\barD_k \left(\barn^{(i} \Delta n^{k)}\right)\, .
\label{midstep}
\end{eqnarray}
We have seen before that $\barn_i$ and $n_i$ are proportional (as both 
are proportional to $\partial_i s$) and so $\barn_i \Delta \sigma^{ik}
= 0$. Hence, we may freely shift the derivative $\barD_k$ off of
$\Delta\sigma^{ik}$ and onto $\barn_i$. Next, for the last term on
the {\sc rhs} of Eq.~(\ref{midstep}), we use $\barn_i \barD_k \barn^i
= 0$ (following from the normalization of $\barn^i$) and find
\begin{eqnarray}
\half \barn^i h_{il}\barD_k\Delta q^{kl}
&=&   \half \barn^i \Delta h_{il}\barD_k\Delta q^{kl}
    - \half\Delta\sigma^{ik}\barD_k \barn_i
\nonumber \\
& & + \half \barn_{i}\barD_k \left( \Delta n^i \Delta n^k\right)
    + \barn_{k}\barD_i \left(\barn^{[i} \Delta n^{k]}\right)
    + \barD_k \Delta n^k\, .
\label{laststep}
\end{eqnarray}
Finally, plugging (\ref{laststep}) into (\ref{step}), we have the
first identity (\ref{identity1}).

To obtain, the second identity (\ref{identity2}), first simply
multiply the first identity (\ref{identity1}) by the smearing
function $N$, thereby obtaining a new equation which has
$N\barn_k \barD_l (\barn^{[l} \Delta n^{k]})$ as one of its {\sc rhs}
terms. Let us reexpress this term to get the result. Straightaway,
\begin{equation}
N\barn_k \barD_l \left(\barn^{[l} \Delta n^{k]}\right) =
\barn_k \barD_l \left(N \barn^{[l} \Delta n^{k]}\right) +
\half\barn_k\left(\barn^k\Delta n^l - \barn^l \Delta n^k\right)
\barD_l N\, ,
\end{equation}
and into this equation we twice substitute (\ref{idsix}), thereby
reaching
\begin{eqnarray}
N\barn_k \barD_l \left(\barn^{[l} \Delta n^{k]}\right) & = &
\barn_k \barD_l \left(N \barn^{[l} \Delta n^{k]}\right) +
\half
\left[
\barn_i (\Delta q^{il} - \Delta\sigma^{il})
- \barn^l \barn_i\barn_k (\Delta  q^{ik} - \Delta\sigma^{ik})
\right. \nonumber \\
& & \left. + \barn^l \left(\barn_i\Delta n^i\right){}^2
- \barn_i \Delta n^i \Delta n^l
\right]\barD_l N\, .
\end{eqnarray}
Next, appealing to identity (\ref{idfive}), we find
\begin{eqnarray}
N\barn_k \barD_l \left(\barn^{[l} \Delta n^{k]}\right) & = &
\barn_k \barD_l \left(N \barn^{[l} \Delta n^{k]}\right) +
\half
\left[
\barn^{k} \barh_{ik} \Delta  q^{il}
- \barn^l \barh_{ik} (\Delta  q^{ik} - \Delta \sigma^{ik})
\right.  \nonumber \\
& & \left. + \barn^l \barh_{ik}\Delta n^i \Delta n^k
           - \barn_i \Delta n^i \Delta n^l\right]\barD_l N\, .
\end{eqnarray}
The last equation transforms exactly into
\begin{eqnarray}
N\barn_k \barD_l \left(\barn^{[l} \Delta n^{k]}\right) & = &
\barn_k \barD_l \left(N \barn^{[l} \Delta n^{k]}\right) -
\half
\left(
\barn^{k} h^{il} \Delta h_{ik} - \barn^l h^{ik}\Delta h_{ik}
\right)\barD_l N  \nonumber \\
& & + \half\left(\barn^l \barh_{ik}\Delta n^i \Delta n^k
           - \barn_i \Delta n^i \Delta n^l\right)\barD_l N
           + \half \barn^l \barh_{ik} \Delta\sigma^{ik} 
             \barD_l N\, ,
\end{eqnarray}
and using this result in the new equation obtained from
(\ref{identity1}) via multiplication by $N$ we get the
desired identity (\ref{identity2}).

\subsection{Asymptotic solution to the embedding equations}

We now show how to formally solve the system (\ref{system-r0}), 
although our discussion also pertains to the system (\ref{system-r0e}) 
as well as a third system at the next order. As we pointed out in 
{\em Section III.C}, the three systems are formally the same. Let us 
rewrite (\ref{system-r0}), eliminating $f$ and making the substitution 
$g = \sin\theta\tilde{g}$, as  
\begin{subequations}\label{system-r0-b}
\begin{eqnarray}
     \sin\theta 
     \frac{\partial\tilde{g}}{\partial\theta} 
   - \frac{\partial h}{\partial\phi}                
     & = & \alpha - \beta\, , 
     \label{system-r0-b1}                              \\
     \sin\theta 
     \frac{\partial\tilde{g}}{\partial\phi} 
   + \sin^2\theta \frac{\partial h}{\partial\theta} & = & \gamma\, .
     \label{system-r0-b2}
\end{eqnarray}
\end{subequations}
Applying $\partial/\partial\theta$ and $(1/\sin^2\theta)
\partial/\partial\phi$ to the 
first and the second lines of (\ref{system-r0-b}) respectively, we 
find upon elimination of $h$ that
\begin{equation}
   \left[ \frac{1}{\sin\theta} \frac{\partial}{\partial\theta}
   \left(\sin\theta
   \frac{\partial}{\partial\theta}\right) 
  +\frac{1}{\sin^2\theta}
   \frac{\partial^2}{\partial\phi^2}\right]\tilde g 
  =\psi(\theta,\phi)
\label{rhs-zeta}
\end{equation}
with
\begin{equation}
   \psi = \frac{1}{\sin\theta}
   \left(
   \frac{\partial\alpha}{\partial\theta}
  -\frac{\partial\beta}{\partial\theta}
   \right) 
  +\frac{1}{\sin^3\theta}
   \frac{\partial\gamma}{\partial\phi}\, .
\label{zeta-def}
\end{equation}
Whence we have transformed the system (\ref{system-r0}) to (\ref{rhs-zeta}),
that is to say the Poisson equation
\begin{equation}
   \Delta \tilde g = \psi\, ,
   \label{poisson-zeta}
\end{equation}
where now and from now on $\Delta$ 
is the Laplacian on the unit sphere. 
We may similarly establish that
\begin{equation}
\Delta h = \chi\, ,
\end{equation}
with
\begin{equation}
   \chi =\frac{1}{\sin^2\theta}
   \left(
   \frac{\partial\beta}{\partial\phi}
  -\frac{\partial\alpha}{\partial\phi}
   \right) 
  +\frac{1}{\sin\theta}
   \frac{\partial}{\partial\theta}
   \left(\frac{\gamma}{\sin\theta}
   \right)\, .
\end{equation}
Solvability of these $S^2$ Poisson equations is ensured if the
following compatibility conditions are satisfied:
\begin{equation}
\oint d\Omega\psi = 0 = \oint d\Omega \chi\, .
\label{compatibilitycondition}
\end{equation}
These conditions simply state that the sources $\psi$ and 
$\chi$ lie in the range of the Laplacian. A careful 
treatment of this issue would relate these compatibility conditions 
to the regularity of $\alpha$, $\beta$, and $\gamma$. 
Indeed, in the axially symmetric case these conditions are tantamount
to the statement that the $B$ metric $\sigma_{ab}$ is free of conical 
singularities at the north and south pole at leading and 
next--to--leading order. In the general case we expect that the 
conditions (\ref{compatibilitycondition}) are also related to the
absence of canonical singularities at the poles. Recall that our  
discussion here is also meant to address the system  (\ref{system-r0e})
as well as a third system at the next order. Such 
compatibility conditions will also crop up when examining 
these systems. We shall require whatever conditions [on
the coefficients ${\cal A}$, ${\cal B}$, and ${\cal G}$
for the system (\ref{system-r0e}) and on similar coefficients
for the next order] are necessary in order that these systems
are solvable.

A solution to (\ref{poisson-zeta}) is formally a solution to the 
original system (\ref{system-r0-b}). To verify that this is indeed the 
case, integrate Eq.~(\ref{system-r0-b1}) to reach
\begin{equation}
   h = \int d\phi
       \left(\sin\theta 
       \frac{\partial\tilde{g}}{\partial\theta} +
       \beta-\alpha 
       \right)\, .
\end{equation}
Substitution of this equation into Eq.~(\ref{system-r0-b2})
then yields
\begin{equation}
   \int d\phi\sin^3\theta 
   \left(\Delta \tilde g - \psi
   \right) = 0\, ,
\end{equation} 
which clearly holds for solutions to (\ref{poisson-zeta}).

We solve (\ref{poisson-zeta}) via standard methods. We expand $\psi$
with respect to spherical harmonics:
\begin{equation}
\psi = \sum_{l,m}\psi_{lm} Y_{lm}(\theta,\phi)\, ,
\label{zeta}
\end{equation} 
and are assuming that as a function $\psi$ at least lies in the
space $L_2(S^2,d\Omega)$ of square integrable functions. But this is 
certainly the case, as we expect that the metric coefficients $\alpha$,
$\beta$, and $\gamma$ are $C^2(S^2)$. As discussed below, $\psi$ is
of even parity and, moreover, must satisfy (\ref{compatibilitycondition}). 
Therefore, the summation in (\ref{zeta}) is over even values of $l$
and with $\psi_{00} = 0$ as the constant mode so that $l$ is never zero in
the sum. 

Search for a solution taking the form 
\begin{equation}
 \tilde{g} =\sum_{l,m}\tilde g_{lm} Y_{lm}(\theta,\phi)\, .
\label{tilde-g}
\end{equation}
Substitution of (\ref{zeta}) and (\ref{tilde-g}) into (\ref{poisson-zeta})
determines the sought--for solution as
\begin{equation}
   \tilde g = -\sum_{l,m}\frac{\psi_{lm}}{l(l+1)}
              Y_{lm}(\theta,\phi)\, ,
\label{newgtil}
\end{equation}
since, of course,  $\Delta Y_{lm} = -l(l+1) Y_{lm}$. Note that in 
this equation and the ones to follow, we never divide by zero, since
the constant mode $\psi_{00}$ vanishes. Similar steps lead to 
the expansion
\begin{equation}
   h = -\sum_{l,m}\frac{\chi_{lm}}{l(l+1)}
      Y_{lm}(\theta,\phi)\, .
\end{equation}
From Eq.~(\ref{newgtil})
\begin{equation}
   g = -\sin\theta\sum_{l,m}\frac{\psi_{lm}}{l(l+1)}
      Y_{lm}(\theta,\phi)\, .
\end{equation}
Whence (\ref{system-r0}) gives the desired final result,
\begin{equation}
   f =\alpha 
     +\sum_{l,m}\frac{\psi_{lm}}{l(l+1)}
      \left[
      \cos\theta Y_{lm}(\theta,\phi)
     +\sin\theta \frac{\partial}{\partial\theta}Y_{lm}
      (\theta,\phi)
      \right]\, .
\end{equation}
The solution must not to destroy the approximation scheme.
We expect a good solution provided that the metric coefficients 
$\alpha$, $\beta$, and $\gamma$ both are $C^2(S^2)$ and lead 
only to small corrections ${}^3{\cal R}r^{-3}$ to the leading 
$2/r^2$ behavior for the $B$ Ricci scalar $\cal R$.

Let us now consider the action of the parity operator ${\cal P}$. 
Recalling its defining action ${\cal P}(x^k) = -x^k$, one can show that
${\cal P}(r\, ,\theta\, ,\phi) = (r\, ,\pi -\theta\, ,\phi+\pi)$ 
and that in (\ref{BOM-B}) the terms $\alpha$ and $\beta$ are of the 
even parity while $\gamma$ is of odd parity. Further, we can state 
that $\cos\theta$, $\partial/\partial\phi$, 
$\psi$ from (\ref{zeta-def}), and 
$\Delta$ are of even parity, while $\sin \theta$ and 
$\partial /\partial\theta$ are of odd parity. Then Eq.~(\ref{poisson-zeta}) 
determines ${\cal P}(\tilde{g}) = \tilde{g}$, and, consequently, $g$ is
of odd parity. In tandem with (\ref{system-r0}) this shows that $f$ 
and $h$ are of even parity. We now re--express the transformations 
(\ref{RThetaPhi}) in Cartesian rather than spherical--polar form, 
adopting standard notation $x^1 = x$, $x^2 = y$, $x^3 =  z$, and 
likewise for the $X^k$:
\begin{subequations} \label{XYZ}
\begin{eqnarray}
    X & = & x + r^{-1}\left[xf - yh 
              + \frac{xzg}{\sqrt{x^2 + y^2}}\right]
              + O(r^{-\varepsilon})
                \, ,\\
    Y & = & y + r^{-1}\left[yf + xh 
              + \frac{yzg}{\sqrt{x^2 + y^2}}\right]
              + O(r^{-\varepsilon})
                \, ,\\
    Z & = & z + r^{-1}\left[zf - g\sqrt{x^2 + y^2}\right]
              + O(r^{-\varepsilon})\, .
\end{eqnarray}
\end{subequations}
By inspection we see that all middle terms involving square brackets 
are $O(r^0)$ and of odd parity. Therefore, the transformation 
(\ref{XYZ}) may indeed be rewritten in the form $X^k = 
x^k + {}^{0}\xi^k + O(r^{-\varepsilon})$ with
${}^{0}\xi^k$ an $O(1)$ odd parity function of the angular variables.
Via similar analysis we can in principle solve the other two systems
we have mentioned [one (\ref{system-r0e}) at $O(r^{-1-\varepsilon})$ 
involving $\cal A$, $\cal B$, and $\cal G$, and the other ---not written 
down--- at $O(r^{-2})$]. Whence, in more detail we find that 
(\ref{expansion-X}) holds.

\section{$\Bderiv$ operator and frame and connection}
\subsection{$\Bderiv$ operator}
Here we develop some standard formulae (drawing mostly from 
Refs.~\cite{Walker,PenroseRindler}) necessary for and in the 
notation of this paper.

\subsubsection{General two--surfaces}
Suppose that $B$ is a Riemannian two--manifold equipped with metric 
$\sigma_{ab}$ and compatible covariant derivative operator $d_{a}$. 
Let $m^{a}$, with complex conjugate $\bar{m}{}^{a}$, be a complex null 
vector field on $B$, chosen such that $m_a\bar{m}{}^a = 1$. Then the
set $\{e_{\hat{3}} = m,e_{\hat{4}} = \bar{m}\}$ is a complex null 
dyad. (We use $\hat{3}$ and $\hat{4}$ rather than $\hat{1}$ and $\hat{2}$ 
for the name indices, as we might assume our null dyad completes a 
Newman--Penrose null tetrad. \cite{NewmanPenrose} Were we to work with 
such a spacetime tetrad, we would respectively use
$e_{\hat{1}}$ and $e_{\hat{2}}$ for the outgoing and ingoing (real)
null normals to $B$.) Hatted indices 
$\hat{a},\hat{b},\cdots$ are null 
dyad indices, whereas $a,b,\cdots$ are general frame indices. The 
complex connection coefficient 
\begin{equation}
\omega \equiv \omega_{\hat{4}\hat{3}\hat{3}} = e_{\hat{4}}{}^{a} 
e_{\hat{3}}{}^{c} d_{c} e_{\hat{3}}{}_{a} = \bar{m}{}^a m^c d_c m_a
\label{omega}
\end{equation}
is the only one which need be considered. Note the the coefficients
$\omega_{\hat{a}\hat{b}\hat{c}}$ are not the connection coefficients
$\Gamma_{abc}$ (which are Christoffel symbols if $a,b\cdots$ are
coordinate indices). Introduce the operator 
$\delta \equiv m^a d_a$. From Eq.~(\ref{omega}) we then have
\begin{subequations} \label{deltam}
\begin{eqnarray}
\delta m^{a} & = & \omega m^{a}
\\
\delta\bar{m}^{a} & = & - \omega \bar{m}^{a}
{\,} .
\end{eqnarray}
\end{subequations}
Also introduce a dyad derivative operator $\hat{d}_{a}$ which ``sees'' 
dyad indices, and whose action on a dyad vector $v^{\hat{a}}$ is 
defined as
\begin{equation}
\hat{d}{}_{a} v^{\hat{c}} 
= e_{a}[v^{\hat{c}}] + 
v^{\hat{b}}\omega^{\hat{c}}{}_{\hat{b}a}\, ,
\end{equation}
where $e_a = \partial_a$ if $a,b\cdots$ are coordinate indices.
The operator $m^a \hat{d}_{a}$ is known as $\Bderiv$ in the 
compacted spin coefficient formalism. \cite{PenroseRindler}

Consider a scalar defined on $B$ as follows:
\begin{equation}
\eta \equiv m^{a_{1}} \cdots m^{a_{p + s}} \bar{m}^{c_{1}} \cdots
\bar{m}^{c_{p}} T_{a_{1} \cdots a_{p+s}c_{1} \cdots c_{p}} 
{\,} ,
\label{eta}
\end{equation}
where $T_{a_{1} \cdots a_{p+s}c_{1} \cdots c_{p}}$ is a rank--$(2p + s)$
tensor field on $B$ (note that $s$ may be negative). Under the mapping
\begin{equation}
m^{a} \mapsto e^{{\rm i}\psi} m^{a}
{\,} ,
\end{equation}
$\eta$ behaves as
\begin{equation}
\eta \mapsto e^{{\rm i}s \psi} \eta
{\,} .
\end{equation}
We say that $\eta$ has {\em spin--weight} $s$, or symbolically 
{\sc sw}$(\eta) = s$. Notice that $\eta$ is just a component 
\begin{equation}
T_{
\underbrace{\scriptstyle \hat{3}\cdots \hat{3}}_{p+s}
\underbrace{\scriptstyle \hat{4}\cdots \hat{4}}_{p}
}
\end{equation}
of the tensor $T_{a b \cdots d}$ with respect to 
the null dyad. More generally, we may define a scalar of spin--weight 
$s$ by taking any rank--$q$ covariant $B$ tensor and contracting any
$\half (q+s)$ of its indices on $m^a$ and the remaining $\half (q-s)$ 
indices on $\bar{m}{}^{a}$. For example, $T_{\hat{3}\hat{4}\hat{3}} = 
m^a \bar{m}^b m^c T_{abc}$ defines a spin--weight $1$ scalar. Of course 
for our construction here $\half(q+s)$ should be positive and an 
integer. For example, with the tensor $T_{abc}$
one can not obtain a component 
of spin--weight $2$ or $-2$ (components of spin--weight $-3$, $-1$, $1$,
and $3$ are possible). Just for the sake of concreteness, let us continue
the development of the formalism with a tensor $T_{a b \cdots d}$ 
having the specific index structure given in Eq.~(\ref{eta}). 

Now define the ``eth'' and ``eth--bar'' operators,
\begin{subequations} \label{eth}
\begin{eqnarray}
\Bderiv \eta & \equiv & m^{a_{1}} \cdots m^{a_{p+s}} \bar{m}^{c_{1}}
\cdots \bar{m}^{c_{p}} m^{b} d_{b} 
T_{a_{1} \cdots a_{p+s}c_{1} \cdots c_{p}}
\\
\bar{\Bderiv} \eta & \equiv & m^{a_{1}} \cdots m^{a_{p+s}} \bar{m}^{c_{1}}
\cdots \bar{m}^{c_{p}} \bar{m}^{b} d_{b} 
T_{a_{1} \cdots a_{p+s}c_{1} \cdots c_{p}}
{\,} .
\end{eqnarray}
\end{subequations}
By inspection we see that {\sc sw}$(\Bderiv \eta) = s + 1$ and
{\sc sw}$(\bar{\Bderiv}\eta) = s - 1$. Quick calculations using integration
by parts followed by appeals to Eqs.~(\ref{deltam}) and their complex 
conjugates show that
\begin{subequations}
\begin{eqnarray}
\Bderiv \eta & = & \delta \eta - s\omega\eta
\\
\bar{\Bderiv} \eta & = &
\bar{\delta} \eta + s\bar{\omega}\eta
{\,} .
\end{eqnarray}
\end{subequations}
Moreover, the first equation in (\ref{eth}), for example, may also be 
written as
\begin{equation}
\eth\eta = m^a \hat{d}_a T_{
\underbrace{\scriptstyle \hat{3}\cdots \hat{3}}_{p+s}
\underbrace{\scriptstyle \hat{4}\cdots \hat{4}}_{p}
}\, .
\end{equation}
{\sc Lemma.} The commutator of $\Bderiv$ and $\bar{\Bderiv}$ is
\begin{equation}
(\bar{\Bderiv}\Bderiv - \Bderiv \bar{\Bderiv})\eta 
= {\textstyle \frac{1}{2}}
s {\cal R} \eta {\,} ,
\label{commutator}
\end{equation}
where ${\cal R}$ is the Ricci scalar of $B$. The lemma is nothing more 
than the standard ``Ricci identity'' obeyed by covariant derivative
operators. In lieu of a proof, we establish the identity for a
particular illustrative example. With the tensor $T_{abc}$, build
$\eta = T_{\hat{3}\hat{4}\hat{3}}$, a scalar of spin--weight $1$. Write 
$\Bderiv = m^a\hat{d}_a = \hat{d}_{\hat{3}}$ and similarly $\bar{\Bderiv} = 
\bar{m}^a\hat{d}_a = \hat{d}_{\hat{4}}$. Then we 
have
\begin{eqnarray}
(\bar{\Bderiv}\Bderiv - \Bderiv \bar{\Bderiv})\eta & = &
- 2\hat{d}_{[\hat{3}} \hat{d}_{\hat{4}]} T_{\hat{3}\hat{4}\hat{3}}
\nonumber \\
& = & {\cal R}^{\hat{c}}{}_{\hat{3}\hat{3}\hat{4}} 
  T_{\hat{c}\hat{4}\hat{3}}
+ {\cal R}^{\hat{c}}{}_{\hat{4}\hat{3}\hat{4}} T_{\hat{3}\hat{c}\hat{3}}
+ {\cal R}^{\hat{c}}{}_{\hat{3}\hat{3}\hat{4}} T_{\hat{3}\hat{4}\hat{c}}
\nonumber \\
& = & {\cal R}^{\hat{3}}{}_{\hat{3}\hat{3}\hat{4}} 
  T_{\hat{3}\hat{4}\hat{3}}
+ {\cal R}^{\hat{4}}{}_{\hat{4}\hat{3}\hat{4}} T_{\hat{3}\hat{4}\hat{3}}
+ {\cal R}^{\hat{3}}{}_{\hat{3}\hat{3}\hat{4}} T_{\hat{3}\hat{4}\hat{3}}
\nonumber \\
& = & {\cal R}_{\hat{4}\hat{3}\hat{3}\hat{4}} T_{\hat{3}\hat{4}\hat{3}}
+ {\cal R}_{\hat{3}\hat{4}\hat{3}\hat{4}} T_{\hat{3}\hat{4}\hat{3}}
+ {\cal R}_{\hat{4}\hat{3}\hat{3}\hat{4}} T_{\hat{3}\hat{4}\hat{3}}\, ,
\end{eqnarray}
where ${\cal R}_{\hat{a}\hat{b}\hat{c}\hat{d}}$ are the dyad components of
the $B$ Riemann tensor. The dyad curvature 
${\cal R}_{\hat{a}\hat{b}\hat{c}\hat{d}}$ is antisymmetric in its first
(and last) pair of indices, and the Ricci scalar is ${\cal R} = 
- 2 {\cal R}_{\hat{3}\hat{4}\hat{3}\hat{4}}$. Hence, we indeed have 
$2[\bar{\Bderiv},\Bderiv]\eta = \frac{1}{2} {\cal R} \eta$.

\subsubsection{Round spheres: moving frame and coordinates}
Let us consider a round sphere of radius $r$ 
sitting in Euclidean three-space $E^3$. To highlight the fact
that we are now working with a round sphere, let us use 
$\mu$ and $\bar{\mu}$ in place of $m$ and $\bar{m}$ for legs
of the complex dyad. On $E^3$ we choose the 
moving frame $\{e_{\vdash},e_{\hat{3}},e_{\hat{4}}\} = 
\{\nu,\mu,\bar{\mu}\}$, with $\nu = \partial/\partial r$ and 
the complex leg
\begin{equation}
\mu^{i}\partial/\partial x^{i} = - \sqrt{\textstyle \frac{1}{2}}
r^{-1} e^{-{\rm i}\phi}[\partial/\partial\theta + {\rm i}
(\sin\theta)^{-1} \partial/\partial\phi]
 = \sqrt{\textstyle \frac{1}{2}} r^{-1}P\partial/\partial\zeta
{\,} .
\label{tangent}
\end{equation}
Here $\zeta = e^{{\rm i}\phi} \cot(\theta/2)$ is the stereographic
coordinate and $P = 1 + \zeta \bar{\zeta}$. These conventions 
agree with Dougan's. \cite{Dougan} Clearly, $\mu^i$ points
everywhere tangent to the foliation of $E^3$ into level--$r$ spheres.
Restriction of $\mu^i$ to a particular level--$r$ sphere determines 
a $\mu^a$ as before (but before denoted $m^a$). With this choice of 
$\mu^{a}$ we find the following connection coefficient
\begin{equation}
\omega \equiv \omega_{\hat{4}\hat{3}\hat{3}} = 
- \sqrt{\textstyle \frac{1}{2}}\bar{\zeta}/r
{\,} .
\end{equation}
When $B$ is the unit sphere $S^2$, we use the notation
$\omega_0$ for this connection coefficient and also 
sometimes $\alpha^{0} = 
\sqrt{\frac{1}{8}}\zeta$ for $-\frac{1}{2}\bar{\omega}_0$.
Here the notation is seemingly odd, but we may adopt it to have 
agreement with the standard Newman--Penrose formalism.
\cite{NewmanPenrose,Dougan}
Adopting this notation, we may consistently define
\begin{subequations} \label{sphericaleth}
\begin{eqnarray}
\Bderiv_0 \eta & \equiv & (\delta_0 + 
2 s \bar{\alpha}^{0}) \eta
\\
\bar{\Bderiv}_0 \eta & \equiv & (\bar{\delta}_0 - 
2 s \alpha^{0}) \eta
{\,} ,
\end{eqnarray} 
\end{subequations}
where $\delta_0 = \sqrt{\frac{1}{2}} P \partial/\partial\zeta$.
Now the commutator considered in the lemma above is
\begin{equation}
(\bar{\Bderiv}_0\Bderiv_0 - \Bderiv_0 \bar{\Bderiv}_0)\eta = 
s\eta {\,} ,
\label{unitcommutator}
\end{equation}
since the Ricci scalar for the unit sphere is the constant function 
$2$. Two other useful round--sphere formulae are 
$\oint d\Omega\, \chi \Bderiv_0 \eta
= -\oint d\Omega\, \eta \Bderiv_0 \chi$, where {\sc sw}$(\chi)$ + {\sc 
sw}$(\eta) = -1$, and $\partial_k V^k = 
(2V_\nu + \Bderiv_0 V_{\bar{\mu}} + \bar{\Bderiv}_0 V_\mu) r^{-1}
+ \partial V_\nu/\partial r$, where $V^k$ is some Cartesian vector.

\subsubsection{Spin--$0$ spherical harmonics}
Let us now document some standard results concerning the unit sphere
Laplacian $\Delta$. First, when acting on scalars of zero 
spin--weight, 
\begin{equation}
2\bar{\Bderiv}_0\Bderiv_0 = \Delta \equiv \frac{1}{\sin\theta}
\frac{\partial}{\partial\theta}\left[
\sin\theta\frac{\partial}{\partial\theta}\right] 
+ \frac{1}{\sin^2\theta}
\frac{\partial^2}{\partial\phi^2}\, .
\end{equation}
The eigenfunctions of $\Delta$ are the spherical harmonics 
$Y_{lm}(\theta,\phi)$ ($m$ runs from $-l$ to $l$ in integer steps), 
which we may also view as functions 
$Y_{lm}(\zeta,\bar{\zeta})$. These obey {\sc sw}$(Y_{lm}) = 0$ 
and $\Delta Y_{lm} = -l(l+1) Y_{lm}$. We may realize 
the spherical harmonics with the following construction.

Consider the standard stereographic projection. \cite{Conway} 
Namely, given a complex number $\zeta$, we produce a unit vector 
$\nu^i = (\nu^1,\nu^2,\nu^3)$ via the formulae
\begin{equation}
\nu^1 P = \zeta + \bar{\zeta}\, ,\qquad
\nu^2 P = -{\rm i}(\zeta - \bar{\zeta})\, ,\qquad
\nu^3 P = P - 2\, ,
\end{equation}
whence we indeed have $\delta_{ij} \nu^i \nu^j = 1$. In these and 
following formulae, $i,j,\cdots$ are $E^3$ Cartesian coordinate
indices. Viewing $\zeta$ as a 
complex coordinate on the unit sphere, we may then produce a Cartesian
point $x^i = r(\nu^1,\nu^2,\nu^3)$ given ``polar coordinates'' 
$(r,\zeta)$. The Cartesian components,
\begin{equation}
\mu^i =  \sqrt{\textstyle{\frac{1}{2}}}r^{-1}P\partial 
x^i/\partial\zeta
= \sqrt{\textstyle{\frac{1}{2}}}P^{-1}\Big(1-\bar{\zeta}{}^2,
-{\rm i}\big(1+\bar{\zeta}{}^2\big), 2\bar{\zeta}\Big)\, ,
\end{equation}
of the vector (\ref{tangent}) are obtained
via the chain rule. Now choose an arbitrary {\em constant} 
Cartesian vector field $c^i$. Clearly then {\sc sw}$(\mu^i c_i) = 1$ 
and {\sc sw}$(\nu^i c_i) = 0$. It follows that, viewed as 
functions on $S^2$, 
the components $\mu^i$ and $\nu^i$ have spin--weights $1$ and $0$
respectively. Therefore, computations based on the 
rules (\ref{sphericaleth}) give the useful formulae
\begin{equation}
\Bderiv_0 \nu^i = \mu^i\, ,\qquad
\Bderiv_0 \mu^i = 0\, ,\qquad 
\bar{\Bderiv}_0 \mu^i = - \nu^i\, .
\label{ethonmnm}
\end{equation}
With these formulae, it is easy to verify that 
$2\bar{\Bderiv}_0\Bderiv_0 \nu^i = -2 \nu^i$; that is to say, the 
$\nu^i$ are essentially the $Y_{1m}$. Moreover, defining
$f^{ij} = \nu^i \nu^j + \mu^i\bar{\mu}^j + \bar{\mu}^i\mu^j$,
one can check with (\ref{ethonmnm}) that
\begin{equation}
2\bar{\Bderiv}_0\Bderiv_0 (\nu^i\nu^j 
  - {\textstyle \frac{1}{3}} f^{ij})
= -6 (\nu^i\nu^j - {\textstyle \frac{1}{3}} f^{ij}) \, ,
\end{equation} 
thereby showing the $\nu^i\nu^j - {\textstyle \frac{1}{3}} f^{ij}$ to 
be essentially the $Y_{2m}$. Naively, there are six 
$\nu^i\nu^j - {\textstyle \frac{1}{3}} f^{ij}$, but only 
five $Y_{2m}$. However, notice that 
$\nu^i\nu_i - {\textstyle \frac{1}{3}} f^{i}{}_{i} = 0$ 
(one condition); hence, $\nu^i\nu^j - {\textstyle \frac{1}{3}} f^{ij}$ 
has only five independent components as expected. As is well--known,
one may continue building higher zero spin--weight harmonics as 
symmetric trace--free Cartesian tensors polynomial in $\nu^i$ 
and $f^{ij}$.

\subsubsection{Spin--$s$ spherical harmonics} 
Extension of the action of
$\Delta \equiv 2\bar{\Bderiv}_0\Bderiv_0$ to scalars of 
arbitrary spin--weight is trivial, since $2\bar{\Bderiv}_0\Bderiv_0$ is 
defined on arbitrary scalars. Eq.~(\ref{unitcommutator}) shows that
on an $s$ spin--weight scalar
\begin{equation}
2\Bderiv_0\bar{\Bderiv}_0 = \Delta - 2s\, .
\label{almostLapl}
\end{equation}

The spin--$s$ spherical harmonics ${}_s Y_{lm}$ are the $s$ spin--weight 
eigenfunctions of the Laplacian $\Delta$. Following Goldberg 
{\em et.~al.}~in Ref.~\cite{Goldberg}, we define
\begin{subequations}\label{spinsharmonics}
\begin{eqnarray}
{}_{s} Y_{lm} & \equiv & [2^s(l-s)!/(l+s)!]^{1/2} \Bderiv_0^s Y_{lm} 
\\
{}_{-s} Y_{lm} & \equiv & [2^{s}(l-s)!/(l+s)!]^{1/2} (-1)^s
\bar{\Bderiv}_0^{s} Y_{lm}\, ,
\end{eqnarray}
\end{subequations}
here with the restriction $0 \leq s \leq l$. 
The discrepancies in factors of $2^{s/2}$ with the definitions in 
Ref.~\cite{Goldberg} stem from the fact that the $\Bderiv_0$ 
operator of Goldberg {\em et.~al.}~is $\sqrt{2}$ times our own.
Notice that we are not allowed to 
increment the spin--weight beyond the 
range $-l$ to $l$ (the same restriction
on the integer $m$). This is suggested by 
the considerations above. Indeed,
the formulae in (\ref{ethonmnm}) clearly 
show that both $\Bderiv_0^2$ and
$\bar{\Bderiv}_0^2$ annihilate $\nu^i$; that is to say, 
$\Bderiv_0 {}_1 Y_{1m}
= 0 = \bar{\Bderiv}_0 {}_{-1}Y_{1m}$. Furthermore, repeated use of these 
formulae shows that both $\Bderiv_0^3$ and 
$\bar{\Bderiv}_0^3$ annihilate 
$\nu^i\nu^j - {\textstyle \frac{1}{3}} f^{ij}$, our geometric 
representation of the $Y_{2m}$.
This then implies $\Bderiv_0 {}_2 Y_{2m} = 
0 = \bar{\Bderiv}_0 {}_{-2}Y_{2m}$. 
These concrete examples exhibit the idea behind the identities
\begin{equation}
\Bderiv_0 {}_l Y_{lm} = 0 = \bar{\Bderiv}_0 {}_{-l}Y_{lm}\, .
\label{topbottomrung}
\end{equation}

The standard identity $\bar{Y}_{lm} = (-1)^m Y_{lm}$ along with the 
definitions of $\Bderiv_0$ and $\bar{\Bderiv}_0$ determines that
\begin{equation}
{}_s\bar{Y}_{lm} = (-1)^{m+s} {}_{-s} Y_{l,-m}\, ,
\label{conjugate}
\end{equation}
now with $-l \leq s \leq l$. We have the main
\\
{\sc Lemma.}  With the Laplacian $\Delta = 
2\bar{\Bderiv}_0\Bderiv_0$ 
and for $-l \leq s \leq l$, 
\begin{equation}
\Delta\, {}_s Y_{lm} = [s(s+1) - l(l+1)] {}_s Y_{lm}\, .
\end{equation}
To prove the lemma, first establish the positive $s$ case via a simple
induction argument, one using Eq.~(\ref{unitcommutator}). Next, obtain
the negative $s$ case with the identities (\ref{conjugate}) and 
(\ref{almostLapl}). 

With the formulae collected so far, we gather the results,
\begin{subequations} 
\begin{eqnarray}
\Bderiv_0 {}_s Y_{lm} & = & [(l-s)(l+s+1)/2]^{1/2} 
            {}_{s+1} Y_{lm} 
\\
\bar{\Bderiv}_0 {}_s Y_{lm} & = & - [(l+s)(l-s+1)/2]^{1/2} 
               {}_{s-1} Y_{lm}\, ,
\end{eqnarray}
\end{subequations}
which augment those given in Eq.~(\ref{topbottomrung}). 
Hence we may view $\Bderiv_0$ as a kind of raising operator and 
$\bar{\Bderiv}_0$ as the corresponding lowering operator.

\subsection{Frame and connection}
The spatial frame 
\begin{subequations}\label{Sigmaframe}
\begin{eqnarray}
e_{\vdash} = \frac{1}{M}\left(\frac{\partial}{\partial r}
- W^a \frac{\partial}{\partial x^a}\right) & = &
\frac{1}{M}\left(\frac{\partial}{\partial r} 
- W_{\bar{m}} m^a \frac{\partial}{\partial x^a}
- W_{m} \bar{m}{}^a \frac{\partial}{\partial x^a}\right)
\label{Sigmaframea}
\\
e_{\hat{3}} = m^a \frac{\partial}{\partial x^a} & = &
\frac{1}{|m_{\bar{\zeta}}|^2 - |m_{\zeta}|^2}
\left(m_{\bar{\zeta}} \frac{\partial}{\partial \zeta}
- m_{\zeta} \frac{\partial}{\partial\bar{\zeta}}\right)
\label{Sigmaframeb}
\end{eqnarray}
\end{subequations}
is dual to (\ref{Sigmacoframe}). Note that $\overline{m_{\zeta}} = 
\bar{m}_{\bar{\zeta}}$ and $\overline{m_{\bar{\zeta}}} = \bar{m}_{\zeta}$.
The basic coframe variables are then seen to be $M$, 
$W_{m}$, $m_{\zeta}$, and $m_{\bar{\zeta}}$.

We may now use the method of Cartan to calculate the connection 
coefficients $\omega_{\hat{\jmath}\hat{k}\hat{l}}$ determined by the 
co--frame (\ref{Sigmacoframe}). The method starts with the no--torsion
formula
\begin{equation}
{\rm d} e^{\hat{k}} = 
- \omega^{\hat{k}}{}_{\hat{l}} \wedge e^{\hat{l}}\, ,
\end{equation}
and the identities 
\begin{subequations}\label{dzetaid}
\begin{eqnarray}
{\rm d}r \wedge {\rm d}\zeta 
& = & \frac{1}{M(|m_{\bar{\zeta}}|^2 - |m_{\zeta}|^2)}
\left(
m_{\bar{\zeta}} e^{\vdash} \wedge e^{\hat{3}} 
- \bar{m}_{\bar{\zeta}} 
e^{\vdash} \wedge e^{\hat{4}}\right) 
\\
{\rm d}\zeta \wedge {\rm d}\bar{\zeta} & = &
\frac{1}{M(|m_{\bar{\zeta}}|^2 - |m_{\zeta}|^2)}
\left(M
e^{\hat{3}} \wedge e^{\hat{4}}
+ W_{m} e^{\vdash} \wedge e^{\hat{3}}
- W_{\bar{m}} e^{\vdash} \wedge e^{\hat{4}}
\right)
\end{eqnarray}
\end{subequations}
prove useful in carrying out the necessary calculations. In any case, 
the list of $\omega_{\hat{\jmath}\hat{k}\hat{l}}$ is the following:
[cf.~Eq.~(\ref{detsigma})]
\begin{subequations}\label{Sigmaconnection}
\begin{eqnarray}
\omega_{\hat{3}\vdash\vdash} & = & -\Bderiv \log M
\\
& & \nonumber \\
\omega_{\hat{4}\hat{3}\hat{3}} & = & 
\frac{1}{|m_{\bar{\zeta}}|^2 - |m_{\zeta}|^2}
\left(\frac{\partial m_{\bar{\zeta}}}{\partial\zeta} 
- \frac{\partial m_{\zeta}}{\partial\bar{\zeta}}\right)
\\
& & \nonumber \\
\omega_{\hat{3}\vdash\hat{4}}& = & \frac{1}{2M}\left[
\frac{(|m_{\bar{\zeta}}|^2 - |m_{\zeta}|^2)'}{
|m_{\bar{\zeta}}|^2 - |m_{\zeta}|^2}
-\Bderiv W_{\bar{m}} - \bar{\Bderiv} W_{m}\right] 
\\
& & \nonumber \\
\omega_{\hat{3}\vdash\hat{3}} & = & 
\frac{1}{M(|m_{\bar{\zeta}}|^2 - |m_{\zeta}|^2)}
\left[m_{\bar{\zeta}} (m_{\zeta})'
- m_{\zeta} (m_{\bar{\zeta}})'\right] 
- \frac{1}{M} \Bderiv W_{m}
\\
& & \nonumber \\
\omega_{\hat{3}\hat{4}\vdash} & =  &
\frac{(m_{\bar{\zeta}})^2(\bar{m}_{\zeta}/m_{\bar{\zeta}})'
- (m_{\zeta})^2(\bar{m}_{\bar{\zeta}}/m_{\zeta})'}{2M
(|m_{\bar{\zeta}}|^2 - |m_{\zeta}|^2)} 
\nonumber \\
& & 
+ \frac{1}{M}\left(\half\bar{\Bderiv}W_{m} 
                  -\half\Bderiv W_{\bar{m}}
+ W_{\bar{m}} \omega_{\hat{4}\hat{3}\hat{3}}
- W_{m} \omega_{\hat{3}\hat{4}\hat{4}}\right)\, ,
\end{eqnarray}
\end{subequations}
where the prime denotes differentiation by $\partial/\partial r$. 
Moreover, with $\omega \equiv \omega_{\hat{4}\hat{3}\hat{3}}$, the
action of $\Bderiv$ is that of $e_{\hat{3}}$, $e_{\hat{3}} - \omega$, 
$e_{\hat{3}} + \omega$, on $\log M$, $W_{m}$, and $W_{\bar{m}}$
respectively (consistent with the defining action of $\Bderiv$ 
given above).

Let us obtain an expression for the scalar curvature ${\cal R}$ of
$B$. Denote by $\theta^{\hat{3}} = \bar{m}_{a} dx^a$ the pullback
of the co--frame $e^{\hat{3}}$ to a $B$ surface, where $d$ is the $B$
exterior derivative (the exterior derivative of $\Sigma$ has appeared
as ${\rm d}$). Then the connection one--form on $B$ is
\begin{equation}
\omega_{\hat{4}\hat{3}\hat{3}}\theta^{\hat{3}} + 
\omega_{\hat{4}\hat{3}\hat{4}}\theta^{\hat{4}} =
\omega \theta^{\hat{3}} - \bar{\omega} \theta^{\hat{4}}\, .
\end{equation}
In terms of the dyad components of the $B$ Riemann tensor, the $B$ 
scalar curvature is ${\cal R} = 
- 2{\cal R}_{\hat{4}\hat{3}\hat{4}\hat{3}}$, whence we may obtain
${\cal R}$ from the formula
\begin{equation}
{\cal R}_{\hat{4}\hat{3}\hat{4}\hat{3}} 
\theta^{\hat{4}}\wedge\theta^{\hat{3}} = 
{\rm Im}\left[d\big(\omega\theta^{\hat{3}}\big)\right]\, .
\end{equation}
We find
\begin{equation}
{\cal R} = -2(e_{\hat{4}}[\omega] + e_{\hat{3}}[\bar{\omega}] 
+ 2\omega\bar{\omega})
\label{Bcurvature}
\end{equation}
as the result.

\section{Divergence of the base Hamiltonian}
Assuming both the fall--off given in Eqs.~(\ref{BOMzero},\ref{BOM}) 
and the center--of--mass scenario with lapse given by Eq.~(\ref{N}), 
in this appendix we examine the ``base Hamiltonian'' $H_\Sigma$, 
isolating its divergent contribution in the $r\rightarrow\infty$ 
limit. We point out that our calculations here complement those given 
in Appendix C. of Ref.~\cite{BeigOMurchadha}, 
which examined the same issue via
a different method. The method we adopt here brings our 
{\em Section IV} results into sharper focus. Moreover, the 
discussion in {\em Section V} is based in part on the results of this 
appendix. We now consider $H_\Sigma$ off--shell; we do not assume that 
the $\Sigma$ initial data obeys the scalar constraint ${\cal H} = 0$. 

\subsection{Geometric identities and two lemmas}
Before turning to $H_\Sigma$, let us first collect some geometric 
identities and prove two lemmas. Consider the following expression 
for the $\Sigma$ scalar curvature:
\begin{equation} 
R = {\cal R} + k^2 - k_{ab} k^{ab}
  + 2 D_j\big(k n^j + b^j\big)\, . 
\label{SigmascalarR} \end{equation} 
In this equation we may write $k^2 - k_{ab}k^{ab} = \half k^2 - 2 k_{mm}
k_{\bar{m}\bar{m}}$ with $k = 2 k_{m\bar{m}}$ as before. Moreover, 
here $b_j = n^k D_k n_j = - \sigma_j{}^k D_k\log M$, where the last 
equality follows from the fact that the $\Sigma$ Levi--Civita connection 
is torsion--free. One way to verify it is to use $n_j = M D_j r$ in
$n^k D_k n_j$ and then carry through with some algebra, along the way 
using the identities $D_k D_j r = D_j D_k r$ (by no torsion) and
$n^k D_j n_k = 0$  (by the normalization of $n^k$). 
Note that $b^j$ points everywhere tangent 
to the $B$ foliation (i.e. $b_j n^j = 0$), whence the single complex 
component $b_{m} = -\Bderiv \log M$ completely specifies this 
$B$ vector. A simple strategy based on the Ricci identity for 
deriving (\ref{SigmascalarR}) is outlined in the appendix of [5]. 
From Eq.~(\ref{SigmascalarR}) we get
\begin{equation} 
R = {\cal R} - {\textstyle \frac{3}{2}} k^2 - 
2 k_{mm} k_{\bar{m}\bar{m}} 
+ 2M^{-1}k'
- 2M^{-1}\big(W_{m}\bar{\Bderiv} k +
W_{\bar{m}}\Bderiv k)
- 4M^{-1}\Bderiv\bar{\Bderiv} M \, .
\label{finalSigscalar}
\end{equation}
To reach this equation we have used the following identities: 
$D_j n^j 
= -k$; $D_j b^j = \sigma^{ij} D_i b_j + n^i n^j D_i b_j = 
(\Bderiv  b_{\bar{m}} + \bar{\Bderiv} b_{m}) - b_a b^a$,
where the term within the parenthesis is the $B$ divergence of the
$B$ vector $b^a$; and
\begin{equation} 
n^k \partial/\partial x^k =
M^{-1}\big(\partial/\partial r 
- W_{m}\bar{\Bderiv} - W_{\bar{m}} \Bderiv\big)\, ,
\end{equation} 
which is just Eq.~(\ref{Sigmaframea}).
In Eq.~(\ref{finalSigscalar}) the terms involving 
$k_{mm}k_{\bar{m}\bar{m}}$ and $W_{m}\bar{\Bderiv} k$ are each
$O(r^{-4})$ and of even parity, as shown by the 
asymptotic expansions 
(\ref{MandWb},\ref{ethexpansion},\ref{expansionfork},\ref{expansionforkmm}) 
and the parity discussion found in 
{\em Section IV.B}. Moreover, 
from Eq.~(\ref{hqMnexpansionsc}) we infer that
\begin{equation}
M^{-1} \sim 1 - {}^1\!M r^{-1} - {}^{(1+\varepsilon)}\!M 
r^{-1-\varepsilon} - \big({}^2\!M + {\rm E.P.T.}\big) r^{-2}\, ,
\end{equation}
where we recall that ${}^1\!M$ is of even parity. These 
considerations and our {\em Section IV} results show that $R$ 
has the expansion $R \sim {}^3R r^{-3} + {}^{(3+\varepsilon)}R
r^{-3-\varepsilon} + {}^4R r^{-4}$ with
\begin{subequations} \label{asympexpforR}
\begin{eqnarray}
{}^3R & = & 2\, {}^2k + {}^3{\cal R} 
- 4(1 + \Bderiv_0\bar{\Bderiv}_0){}^1\!M
\label{asympexpforRa}\\
{}^{(3+\varepsilon)}R
& = & 
2(1-\varepsilon)\,{}^{(2+\varepsilon)}k 
+ {}^{(3+\varepsilon)}{\cal R}
- 4(1 + \Bderiv_0\bar{\Bderiv}_0){}^{(1+\varepsilon)}\!M
\label{asympexpforRb}\\
{}^{4}R & =  & 
{}^4{\cal R}
- 4(1 + \Bderiv_0\bar{\Bderiv}_0){}^2\!M + {\rm E.P.T}\, .
\label{asympexpforRc}
\end{eqnarray}
\end{subequations}
The coefficient ${}^3R$ is easily seen to 
be of even parity via {\em Section IV} results and arguments. 

We now state the lemmas. Let $\beta^{\bot}{}_\nu = 
\beta^\bot{}_i \nu^i$ as before. Then we have the following:

\noindent {\sc Lemma 1.}
\begin{equation}
\oint d\Omega\beta^{\bot}{}_\nu\, {}^{(3+\varepsilon)}R
= (1-\varepsilon)\oint d\Omega\beta^{\bot}{}_\nu\, {}^{(3+\varepsilon)}
\big[\sigma^{\mu\nu}\sigma^{\lambda\kappa}
\Re_{\mu\lambda\nu\kappa}\big]\, .
\end{equation}

\noindent {\sc Lemma 2.}
\begin{equation}
\oint d\Omega\beta^{\bot}{}_\nu\, {}^4R = 0\, .
\end{equation}

\noindent To prove {\sc Lemma} 1, first use 
Eqs.~(\ref{RiemanntoRiemann},\ref{kcalRRiemanna},\ref{asympexpforRb}) 
to obtain 
\begin{eqnarray}
\oint d\Omega\beta^{\bot}{}_\nu\, {}^{(3+\varepsilon)}R
 & = & (1-\varepsilon)
 \oint d\Omega\beta^{\bot}{}_\nu\, {}^{(3+\varepsilon)}
\big[\sigma^{ij}\sigma^{kl}
R_{ikjl}\big] 
\nonumber \\ 
 & & 
+  \oint d\Omega\beta^{\bot}{}_\nu\,
\big[\varepsilon {}^{(3+\varepsilon)}{\cal R} 
- 4(1+\Bderiv_0\bar{\Bderiv}_0){}^{(1+\varepsilon)}\!M\big]\, ,
\end{eqnarray}
and then with Eq.~(\ref{ReRKK}) replace the $\Sigma$ Riemann tensor on the 
{\sc rhs} with the spacetime Riemann tensor. The second integral on the 
{\sc rhs} vanishes. Indeed, double use of angular integration by parts 
along with Eq.~(\ref{ethonmnm}) shows that the 
unit--sphere average of $\beta^\bot{}_\nu(1+\Bderiv_0\bar{\Bderiv}_0)
{}^{(1+\varepsilon)}\!M$ is zero. Moreover,  the unit--sphere
average of $\beta^\bot{}_\nu {}^{(3+\varepsilon)}{\cal R}$ is 
also zero. Indeed, explicitly
\begin{equation}
{}^{(3+\varepsilon)}{\cal R} = \Bderiv^2_0 b_{\bar{\mu}\bar{\mu}}
+ \bar{\Bderiv}{}^2_0 b_{\mu\mu}
- 2\bar{\Bderiv}_0\Bderiv_0 b_{\mu\bar{\mu}}
- 2 b_{\mu\bar{\mu}}\, ,
\label{threeplusecalR}
\end{equation}
that is to say the result (\ref{Rexpansion}) 
only with $a_{ij}$ replaced by $b_{ij}$.
This result can be verified with the techniques of {\em Section IV}.
Angular integration by parts along with (\ref{ethonmnm}) 
again shows that the 
average in question vanishes. (Note, however, that the unit--sphere 
average of ${}^{(3+\varepsilon)}{\cal R}$ need not vanish. The 
discussion in {\em Section V} of the mass--aspect for the energy 
scenario rests on the fact that the same is true for
${}^{3}{\cal R}$.) 
As for {\sc Lemma} 2, the same arguments apply, 
because we expect
${}^{4}{\cal R}$ to have the form
\begin{equation}
{}^{4}{\cal R} = \Bderiv^2_0 c_{\bar{\mu}\bar{\mu}}
+ \bar{\Bderiv}{}^2_0 c_{\mu\mu}
- 2\bar{\Bderiv}_0\Bderiv_0 c_{\mu\bar{\mu}}
- 2 c_{\mu\bar{\mu}} + {\rm E.P.T.}
\end{equation}
This result should follow from Eq.~(\ref{threeplusecalR})
upon setting $\varepsilon = 1$ and including quadratic even--parity 
corrections. 

\subsection{Divergence of the base Hamiltonian}
Let us now recall that $H_\Sigma$ is the volume integral given 
in Eq.~(\ref{smearedHam}), where 
\begin{equation}
{\cal H} = \frac{\sqrt{h}}{16\pi}\big(K_{ij}K^{ij}-K^2-R\big)\, .
\label{Cscalarconstraint}
\end{equation} 
In what follows, let us assume for simplicity that the slice $\Sigma$ is 
topologically Cartesian three--space $R^3$, that we may ignore the issue 
of inner boundary terms. Upon integration against the product of the
$\Sigma$ volume element $d^3x\sqrt{h} = d^2x dr \sqrt{\sigma}M$ and lapse
$N$, the terms in 
(\ref{Cscalarconstraint}) which are quadratic in the $\Sigma$ extrinsic 
curvature tensor $K_{ij}$ do not contribute to the limit. Indeed, these 
terms are $O(r^{-4})$ and of leading even parity, 
the volume element is $O(r^2)$ 
and of leading even parity, 
while $N$ is $O(r)$ and of course of leading odd
parity. Then the product of all of these terms is $O(r^{-1})$ and upon
integration over the radial coordinate would yield a logarithmic
divergence were it not for the fact that this product is of leading odd
parity. 

Let us then focus on the term 
\begin{equation}
-\frac{1}{16\pi}\int_{\Sigma}d^3x\sqrt{h}N R \sim
-\frac{1}{16\pi}\int_{\Sigma}d\Omega dr r^3 \beta^\perp{}_\nu R\, ,
\end{equation} 
where we have used 
Eqs.~(\ref{N},\ref{hqMnexpansionsc},\ref{elementofB}) 
and the parity
properties of leading terms to isolate on the {\sc rhs} only those terms 
from the {\sc lhs} which lead or could lead to an infinite limit. Next, 
we insert the radial expansion for $R$ and integrate in $r$ 
term--by--term, thereby reaching 
\begin{eqnarray}
\lefteqn{-\frac{1}{16\pi}\int_{\Sigma}d^3x\sqrt{h}N R \sim} & & \\ & &
-\frac{r}{16\pi}\oint d\Omega \beta^\perp{}_\nu {}^3R
-\frac{r^{1-\varepsilon}}{16\pi(1-\varepsilon)}\oint d\Omega
\beta^\perp{}_\nu {}^{(3+\varepsilon)}R -\frac{\log r}{16\pi}\oint
d\Omega \beta^\perp{}_\nu {}^4R\, . \nonumber 
\end{eqnarray} 
The first term on the {\sc rhs} vanishes due to the even parity of
${}^3R$. These calculations, the argument of the previous paragraph, 
and the lemmas above then show that 
\begin{equation}
H_\Sigma \sim -\frac{r^{1-\varepsilon}}{16\pi}
\oint d\Omega\beta^{\bot}{}_\nu\, {}^{(3+\varepsilon)}
\big[\sigma^{\mu\nu}\sigma^{\lambda\kappa}   
\Re_{\mu\lambda\nu\kappa}\big] + 0 \cdot \log r\, .
\label{CdivergentH}
\end{equation} 
Note that the coefficient of $\log r$ has been found to vanish both
by parity arguments and {\sc Lemma} 2. 
Because the first divergent term in Eq.~(\ref{CdivergentH}) has 
the opposite sign from the corresponding term in the main text 
[cf.~Eqs.~(\ref{BYintegral},\ref{N},\ref{thirdresult})],
we see that the full Hamiltonian $H_\Sigma + H_B$ is finite even 
off--shell. Whence $H_B$ must be finite on--shell.

Finally, as an aside we note that the results of this appendix show
\begin{equation}   
M^{\perp k}_\infty = - \frac{1}{8\pi}
\oint d\Omega\, {}^4\!\big[n^i n^j R_{ij}\big]\nu^k
\end{equation}
to be equivalent to Eq.~(\ref{prelimMperpk}) as a definition of 
center--of--mass coordinates for initial data sets. This expression 
in tandem with the Gau\ss--Codazzi--Mainardi splitting of the spacetime
Ricci tensor $\Re_{\mu\nu}$ might be used to derive other equivalent  
expressions for center of mass.


\begin{thebibliography}{99}

\bibitem{Schwinger} J.~Schwinger, ``Particles, Sources, and
Fields,'' Vol.~1, p.~24, Addison--Wesley, Reading, 1970.

\bibitem{Weinberg} S.~Weinberg, ``The Quantum Theory of 
Fields,'' Vol.~1, p.~317, Cambridge University Press, Cambridge, 1995.

\bibitem{ReggeTeitelboim} T.~Regge and C.~Teitelboim, 
                          Ann.~Phys.~{\bf 88} (1974), 286.

\bibitem{BeigOMurchadha} R.~Beig and N.~{\sc \'{o}}
                         Murchadha, Ann.~Phys.~{\bf 174} (1987), 
                         463.

\bibitem{BrownYork} J.~D.~Brown and J.~W.~York,
                    Phys.~Rev.~D {\bf 47} (1993), 1407.   

\bibitem{Petrov95} A.~N.~Petrov, Int.~J.~Mod.~Phys.~D {\bf 4} 
                   (1995), 451.

\bibitem{BBWY} See also the earlier work H.~Braden, J.~D.~Brown, 
               B.~Whiting, and J.~W.~York, Phys.~Rev.~D {\bf 42}
               (1990), 3376.

\bibitem{BLYss} J.~D.~Brown, S.~R.~Lau, and J.~W.~York, 
                Phys.~Rev.~D {\bf 59} (1999), 064028. 

\bibitem{Lau2} S.~R.~Lau, Phys.~Rev.~D {\bf 60} (1999), 104034.

\bibitem{ADM} R.~Arnowitt, S.~Deser, and C.~Misner,
              in ``Gravitation: an Introduction 
              to Current Research'' (L.~Witten, Ed.),
              p.~227, Wiley, New York, 1962.

\bibitem{TBS} A.~Trautman, Bull.~Acad.~Pol.~Sci.~S\'{e}rie  
              des Sci.~Math.~Ast.~Phys.~{\bf 6} (1958), 407;
              H.~Bondi, M.~G.~J.~van der Burg, and
              A.~W.~K.~Metzner, Proc.~R.~Soc.~London A {\bf 269}
              (1962), 21; R.~K.~Sachs, Proc.~R.~Soc.~London
              A {\bf 270} (1962), 103; R.~K.~Sachs,
              Phys.~Rev.~{\bf 128} (1962), 2851.


\bibitem{AbbottDeser} L.~Abbott and S.~Deser, Nucl.~Phys.~B
                      {\bf 195} (1982), 76.
                   
\bibitem{HawkingHorowitz} S.~W.~Hawking and G.~T.~Horowitz, 
                          Class.~Quantum Grav.~{\bf 13}
                          (1996), 1487.

\bibitem{BLY1} J.~D.~Brown, S.~R.~Lau, and J.~W.~York,
               Phys.~Rev.~D {\bf 55} (1997), 1977.


\bibitem{Brown} J.~D.~Brown, J.~Creighton, and R.~B.~Mann,
                Phys.~Rev.~D {\bf 50} (1994), 6394.

\bibitem{BLY3} J.~D.~Brown, S.~R.~Lau, and J.~W.~York, 
               Ann.~Phys.~{\bf 297} (2002), 175. 


\bibitem{AshtekarHansen} A.~Ashtekar and R.~O.~Hansen,
                        J.~Math.~Phys.~{\bf 19} (1978), 1542.

\bibitem{Penrose} R.~Penrose, Proc.~Roy.~Soc.~Lond.~A {\bf 381}
                  (1982), 53.

\bibitem{Goldberg0} J.~N.~Goldberg, Phys.~Rev.~D {\bf 41} (1990), 
                    410.

\bibitem{SHayward} S.~A.~Hayward, Phys.~Rev.~D {\bf 49} (1994), 831.

\bibitem{Ashtekar} A.~Ashtekar in ``General Relativity
                   and Gravitation'' (A.~Held, Ed.), 
                   Vol.~2, p.~37, Plenum Press, New York, 1980.

\bibitem{Heinz}  E.~Heinz, J.~Math.~Mech.~{\bf 11} (1962), 421.

\bibitem{Spivak} M.~Spivak, ``A Comprehensive Introduction
                 to Differential Geometry,'' 2nd ed., Vol.~5, 
                 p.~280, Publish or Perish, Inc., Berkeley, 1979.

\bibitem{PP88} A.~D.~Popova and  A.~N.~Petrov, Int.~J.~Mod.~Phys.~A
               {\bf 3} (1988), 2651.

\bibitem{Eisenhart}   L.~P.~Eisenhart, ``Continuous groups
of transformations,'' p.~35, Princeton University Press,
                      Princeton, 1933.

\bibitem{Yorkaction} J.~W.~York in ``Sources of Gravitational
Radiation'' (L.~Smarr, Ed.), p.~83, Cambridge University Press,
Cambridge, 1979; J.~W.~York, Found.~Phys.~{\bf 16} (1986), 249.

\bibitem{ERP} A.~R.~Exton, E.~T.~Newman, and R.~Penrose,
J.~Math.~Phys.~{\bf 10} (1969), 1566.

\bibitem{Tod} K.~P.~Tod, Class.~Quantum Grav.~{\bf 3} (1986), 
1169.

\bibitem{Walker} M.~Walker in ``Gravitational Radiation;
                 NATO Advanced Study Institute, Proceedings of
                 the Les Houches School, 2--21 June 1982''
                 (N.~ Deruelle and T.~Piran, Eds.), p.~145,
                 North Holland, Amsterdam, 1983.

\bibitem{PenroseRindler} R.~Penrose and W.~Rindler, ``Spinors
                         and Spacetime,'' Vol.~1, p.~250, 
                         Cambridge University Press, Cambridge,
                         1984.

\bibitem{NewmanPenrose} See, for example, Ref.~\cite{PenroseRindler}
                    or T.~Newman and K.~P.~Tod in
                    ``General Relativity and
                    Gravitation'' (A.~Held, Ed.), Vol.~2, p.~1,
                    Plenum Press, New York, 1980.

\bibitem{Dougan} A.~J.~Dougan, Class. Quantum Grav. {\bf 9} (1992),
              2461.

\bibitem{Conway} J.~B.~Conway, ``Functions of One Complex Variable,''
                 2nd ed., p.~8, Springer Verlag, New York, 1973.

\bibitem{Goldberg} J.~N.~Goldberg, A.~J.~Macfarlane, E.~T.~Newman,
                   F.~Rohrlich, and E.~C.~G.~Sudarshan,
                   J.~Math.~Phys.~{\bf 8} (1967), 2155.

\bibitem{SzabadosPoinc} L.~B.~Szabados, Class.~Quantum~Grav.~{\bf 20} 
(2003), 2627.

\end{thebibliography}
\end{document}